\documentclass[%
aip,
jcp,%
amsmath,amssymb,
reprint,%
longbibliography,
floatfix,
nofootinbib]{revtex4-2}
\usepackage{graphicx} % Required for inserting images
\usepackage{hyperref}
\hypersetup{colorlinks=true,
	linkcolor=black,
	allcolors=black}
\usepackage[utf8]{inputenc}

\usepackage[outercaption]{sidecap} 
\usepackage{sidecap} 
\sidecaptionvpos{figure}{c}
\usepackage{newtxtext}
\usepackage{newtxmath}
\usepackage{upgreek}
\usepackage[version=4]{mhchem}
\usepackage{physics}
\usepackage{siunitx}

\usepackage{float}
\usepackage[caption = false]{subfig}
\AtBeginDocument{\RenewCommandCopy\qty\SI}
\usepackage{xcolor}

\renewcommand{\transp}{\intercal}
\newcommand{\QM}{\mathrm{QM}}
\newcommand{\MM}{\mathrm{MM}}
\newcommand{\Ef}{\vb*{\mathcal{E}}}
\newcommand{\hlt}[1]{{\color{black}#1}}

\newcommand{\hltt}[1]{{\color{black}#1}}

\usepackage{footmisc}

\begin{document}
	
	\setlength{\abovedisplayskip}{4pt}
	\setlength{\belowdisplayskip}{4pt}
	{
		\makeatletter
		\def\frontmatter@thefootnote{%
			\altaffilletter@sw{\@fnsymbol}{\@fnsymbol}{\csname c@\@mpfn\endcsname}%
		}%
		\makeatother
	
	\title{Efficient polarizable QM/MM using the direct reaction field Hamiltonian with electrostatic potential fitted multipole operators}
	\author{Thomas P. Fay${}^1$}
	\email{tom.patrick.fay@gmail.com}
	\affiliation{${}^1$Aix Marseille Univ, CNRS, ICR, 13397 Marseille, France}
%	\affiliation{	$\text{Corresponding author: tom.patrick.fay@gmail.com}$}
	\author{Nicolas Ferr\'e${}^1$}
	\affiliation{${}^1$Aix Marseille Univ, CNRS, ICR, 13397 Marseille, France}
	\author{Miquel Huix-Rotllant${}^1$}
	\affiliation{${}^1$Aix Marseille Univ, CNRS, ICR, 13397 Marseille, France}

	\begin{abstract}
		
		Electronic polarization and dispersion are decisive actors in determining interaction energies between molecules. These interactions have a particularly profound effect on excitation energies of molecules in complex environments, especially when the excitation involves a significant degree of charge reorganisation. The direct reaction field (DRF) approach, which has seen a recent revival of interest, provides a powerful framework for describing these interactions in quantum mechanics/molecular mechanics (QM/MM) models of systems, where a small subsystem of interest is described using quantum chemical methods and the remainder is treated with a simple MM force field. In this paper we show how the DRF approach can be combined with the electrostatic potential fitted (ESPF) multipole operator description of the QM region charge density, \hlt{which significantly improves the efficiency of the method, particularly for large MM systems, and for typical calculations effectively eliminates the dependence on MM system size}. We also show how the DRF approach can be combined with fluctuating charge descriptions of the polarizable environment, as well as previously used atom-centred dipole-polarizability based models. We further show that the ESPF-DRF method provides an accurate description of molecular interactions in both ground and excited electronic states of the QM system and apply it to predict the gas to aqueous solution solvatochromic shifts in the UV/visible absorption spectrum of acrolein.
		
	\end{abstract}
	
	\maketitle
	
	\section{Introduction}
	
	Many light-activated process in nature and artificial systems occur in complex condensed phase environments. These processes can often be understood in terms of a chemical fragment of importance, containing relatively few electrons and nuclei, embedded in an environment which interacts electrostatically with this fragment. This has motivated the use of Quantum Mechanics/Molecular Mechanics (QM/MM) models\cite{senn_qmmm_2009,tzeliou_review_2022} to study many phenomena in the condensed phase, where electrons in the fragment of interest is treated quantum mechanically, while the rest of the system is treated with a molecular mechanics force field, typically with point charges (and sometimes higher order multipoles\cite{bondanza_polarizable_2020}) describing the electrostatic interactions within the MM system. The complexity in this approach arises in how the interaction between QM and MM subsystems is treated, and the details of this interaction can have profound effects on physically observable properties, particularly those involving excited electronic states of the QM system such as optical absorption spectra\cite{marini_what_2010,nicoli_assessing_2022,giovannini_general_2017,giovannini_quantum_2019,humeniuk_multistate_2024,alias-rodriguez_solvent_2023,loco_modeling_2018} and topologies of conical intersections between excited states.\cite{liu_conical_2022}
	
	It has long been acknowledged that the electronic polarizabity of the environment plays an especially decisive role in determining properties of excited states in condensed phase environments.\cite{bondanza_polarizable_2020} For example electronic polarization of an environment typically stabilises excited states where there is a significant redistribution of charge within a molecular system, which can red shift absorption peaks and have significant effects on electron transfer rates.\cite{marini_what_2010,nicoli_assessing_2022} Likewise changes in dispersion interactions between ground and excited states can influence the optical properties of molecules.\cite{budzak_solvatochromic_2016,renger_theory_2008} Simple fixed charge models for the MM environment cannot capture these effects, and this has driven the development of a range of polarizable QM/MM methods.\cite{marini_what_2010,nicoli_assessing_2022}
	
	Most polarizable QM/MM approaches involve solving the induction equations for dipoles $\vb*{\mu}$ in the MM environment in response to the electric fields $\vb*{\mathcal{E}}$ generated by the MM region and the average charge density of the QM region $\vb*{\mu} = \vb*{\alpha}(\vb*{\mathcal{E}}_\MM + \ev{\vb*{\mathcal{E}}_\QM})$.\cite{bondanza_polarizable_2020,bondanza_openmmpol_2024} The energy is then evaluated using the classical expression $E_\mathrm{pol} = -\frac{1}{2} \vb*{\mu}\cdot (\vb*{\mathcal{E}}_\MM + \ev{\vb*{\mathcal{E}}_\QM}) $. These approaches are termed mean-field (MF) or sometimes self-consistent field (SCF) approaches because they require self-consistent evaluation of the MM induced dipoles and minimisation of the QM/MM energy. This approach has been shown to accurately capture polarization effects, but the extension of the SCF approach to electronic excited states is complicated by the non-linearity in the MF approach.\cite{liu_conical_2022,improta_state-specific_2006} In the state specific approach, this leads to an unphysical breaking of orthogonality between ground and excited state wave functions, as well as numerical issues in converging the coupled dipoles and excited state charge density. Perturbative\cite{slipchenko_solvation_2010} and linear response\cite{lipparini_linear_2012} approaches have been developed to circumvent this issue, but these involve additional approximations which can cause other issues, for example close to conical intersections between electronic excited states.\cite{liu_conical_2022} 
	
	An alternative approach called the Direct Reaction Field (DRF)\cite{thole_quantum_1980,thole_direct_1982} approach has seen a recent revival of interest.\cite{liu_conical_2022,humeniuk_multistate_2024} %, where it has been recently been rigorously formulated with exact electron interaction integrals. %(not to be confused with Discrete Reaction Field, which is a commonly used name for the point dipole polarizable models described above which regrettably shares the same acronym)
	In this approach the MM polarization accounts for instantaneous fluctuations in the QM charge distribution rather than the average charge density in the MF approach. This means that expectation values of the electric field generated by the QM region are simply replaced by the corresponding quantum mechanical operators in the classical polarization energy expressions, $\ev{\vb*{\mathcal{E}}_\QM} \to \hat{\vb*{\mathcal{E}}}_\QM$. The DRF method therefore treats the MM polarization energy as an additional term in the many-body Hamiltonian for the QM subsystem, and thus the MM polarization in response to ground and excited electronic states are treated on equal footing.\cite{liu_conical_2022,humeniuk_multistate_2024} Furthermore since DRF theory features instantaneous electrostatic responses in both QM and MM subsystems dispersion interactions between QM and MM regions are also described by this framework (although the magnitude of dispersion energies are typically overestimated). Original implementations of the DRF method used approximate expansions of the many-body Hamiltonian in terms of Mulliken-type charge operators,\cite{de_vries_implementation_1995} but recently an integral exact formulation of the DRF method (IEDRF)\cite{liu_conical_2022,humeniuk_multistate_2024} has been developed where the DRF Hamiltonian is evaluated treating the additional one- and two-electron interaction integrals exactly. It has been demonstrated that this method captures a range of excited state properties of molecules very accurately,\cite{humeniuk_multistate_2024} but one disadvantage of this approach is the need to know the response of the MM polarization to any possible change in the electric fields at the MM sites, which \hlt{in its most naive form} requires the evaluation and storage of a potentially very large matrix \hlt{MM region polarization response matrix, which we denote $\vb{K}_\mathrm{D}$}. This makes the method \hlt{naively} scale close to $\mathcal{O}((3N_{\MM})^3)$ for very large MM systems, where $N_\mathrm{MM}$ is the number of polarizable sites in the MM region. \hlt{This scaling, and the storage of a large matrix, can be avoided, as pointed out in Ref.~\onlinecite{humeniuk_multistate_2024}, through iterative construction of the IEDRF Hamiltonian, which reduces the scaling to $\mathcal{O}(N_{\mathrm{AO}}^2(3N_{\MM})^2)$ for large MM systems, where $N_{\mathrm{AO}}$ is the number of atomic orbitals used in the description of the QM region.} \hlt{The IEDRF method additionally requires additional non-standard one and two-electron integrals for its implementation in existing electronic structure codes. Although a library for this is now freely available,\cite{humeniuk_efficient_2022} more wide-spread adoption of DRF could be accelerated if these non-standard integrals could be avoided and instead existing integral libraries and quantum chemistry code could be repurposed.}
	
	\hlt{In order to improve the efficiency of the DRF method, particularly for large MM systems, while retaining the strengths of IEDRF,} in this work we propose re-visiting the DRF method but with the electrostatic potential fitted (ESPF) scheme for deriving an atom-centred expansion of the DRF Hamiltonian.\cite{ferre_approximate_2002,huix-rotllant_analytic_2021} Atom-centred ESPF multipole operators have already been proposed as a tool for accelerating standard fixed charge QM/MM calculations, where they facilitate efficient evaluations of electron energies, as well as analytic gradients and hessians of the QM/MM energy.\cite{schwinn_analytic_2019,huix-rotllant_analytic_2021,bonfrate_efficient_2023,bonfrate_analytic_2024} They have also been used to enable the rigorous treatment of periodic boundary conditions in QM/MM calculations. \hltt{Furthermore it has been shown that the ESPF approach provides a sufficiently accurate approximation for mean-field polarizable embedding models.\cite{reinholdt_fast_2022}} In this work we show how ESPF multipole operators can be used to derive an efficient and accurate approximate formulation of the DRF method. \hlt{The ESPF-DRF approach that we propose requires no additional electron integrals beyond those available in all standard libraries, and reduces the scaling of the method from $\mathcal{O}(N_\MM N_\mathrm{AO}^3)$ for IEDRF\cite{humeniuk_multistate_2024} to $\mathcal{O}(N_\QM N_{\mathrm{AO}}^3)$ for typical calculations, and from $\mathcal{O}(N_\MM^2 N_\mathrm{AO}^2)$ to $\mathcal{O}(N_\MM^2 N_{\mathrm{QM}})$ for calculations with very large MM systems.} {The DRF method is also directly compatible with alternatives to point dipole-based models for the MM region polarization \hlt{(for example polarizable continuum methods\cite{de_vries_implementation_1995})} , and we also show how to combine the ESPF-DRF method with the fluctuating charge (FQ) model for the environment polarization (also known as charge equilibration models).\cite{rick_dynamical_1994,lipparini_linear_2012}}
	
	% \begin{enumerate}
		%     \item Polarizability importance for excited states. Solvatochromism. CI structure in condense phase.
		%     \item Mean-field approaches have limitations, particularly for excited states.
		%     \item DRF solves many of these issues. 
		%     \item Integral exact DRF scales poorly with MM system size. Much old Mulliken charge approaches have all the well-known limitations of mulliken charges.
		%     \item ESPF method allows a consistent treatement of electrostatics between QM and MM regions. Has been successfully applied with rigorous PBC QM/MM, to obtain analytic gradients/hessians efficiently, and with MF-pol.
		%     \item Here we demonstrate how to combine DRF with ESPF, and also use the DRF framework with fluctuating charge models. Compatibility with emerging ML models.
		% \end{enumerate}
	\vspace{-10pt}
	\section{Theory}
	\vspace{-10pt}
	In this section we outline the theory of the DRF and ESPF methods, and describe how to combine them. Note that throughout this section we will use $\rho$ to denote a general charge distribution (and not an electron density). Indices $A,B,...$ will be used for nuclei in either the QM or MM regions, $\mu,\nu,...$ will be used as electron atomic-orbital basis indices. $\vb*{r}$ will be used for electron coordinates, $\hat{\vb*{r}}_i$ for the position operator for electron $i$, $\vb*{R}_A$ for QM region nuclear coordinates for atom $A$, and $\vb*{R}_A^{\MM}$ for MM region nuclear coordinates. Atomic units will be assumed throughout.
	\vspace{-10pt}
	\subsection{QM/MM with electrostatic embedding}
	\vspace{-10pt}
	The standard way to treat electrostatic embedding between a sub-system described with some QM method and an MM environment is to start with an energy functional for the whole system of the form\cite{senn_qmmm_2009,tzeliou_review_2022}
	\begin{align}
		E_\mathrm{tot} = E_\QM + E_\MM + E_{\mathrm{int}}
	\end{align}
	where $E_\QM = \ev*{\hat{H}_\QM}$ is the energy of the QM system on its own, and likewise $E_\MM$ is the energy of the bare MM system. The interaction between the QM and MM sub-systems is described by the $E_\mathrm{int}$ term, and when the MM system is described by a fixed charge distribution $\rho_{\MM}(\vb*{r})$ interacting via electrostatic interactions with the QM system, this interaction term is given by
	\begin{align}
		E_\mathrm{int}^{\mathrm{el}} = \int \dd{\vb*{r}} \ev{\rho_\mathrm{QM}(\vb*{r})}\phi_{\mathrm{MM}}(\vb*{r})
	\end{align}
	where $\ev{\rho_\mathrm{QM}(\vb*{r})} = \ev{\rho_\mathrm{QM}^\mathrm{elec}(\vb*{r})}+ \rho_\mathrm{QM}^\mathrm{nuc}(\vb*{r})$ is the charge density of the QM region, where the electronic and nuclear contributions are given by
	\begin{align}
		\ev{\rho_\mathrm{QM}^\mathrm{elec}(\vb*{r})} &= -\sum_{i=1}^{N_\mathrm{\mathrm{e}}}\ev{\delta(\vb*{r}-\hat{\vb*{r}}_i)}{\Psi} \\
		\rho_\mathrm{QM}^\mathrm{nuc}(\vb*{r}) &=\sum_{A=1}^{N_\mathrm{\mathrm{QM}}}Z_A\delta(\vb*{r}-{\vb*{R}}_A)
	\end{align}
	and $\phi_{\mathrm{MM}}(\vb*{r})$ is the electrostatic potential generated by the MM region
	\begin{align}
		\phi_{\mathrm{MM}}(\vb*{r}) = \int\dd{\vb*{r}'} \frac{1}{\|\vb*{r}-\vb*{r}'\|}\rho_{\mathrm{MM}}(\vb*{r}').
	\end{align}
	
	The effects of Pauli repulsion from the MM environment can also be accounted for with general one-body terms of the form
	\begin{align}
		E_\mathrm{int}^{\mathrm{rep}} = \int \dd{\vb*{r}}\int\dd{\vb*{r}'} \Gamma(\vb*{r},\vb*{r}')v_{\mathrm{rep}}(\vb*{r},\vb*{r}').
	\end{align}
	where $\Gamma$ is the spin-free one-body reduced density matrix for the QM system and $v_\mathrm{rep}$ quantifies the Pauli repulsion. Various models have been proposed for $v_\mathrm{rep}$, for example psueudo-potenital models\cite{liu_conical_2022,marefat_khah_avoiding_2020,larsen_does_2010} and the one we use in this work is based on the exchange energy with pseudo-orbitals for the environment,\cite{giovannini_general_2017} as described in Appendix \ref{app-pauli-rep}.
	Both this term and the electrostatic term can be written as an expectation value of a one-electron operator
	\begin{align}
		% E_\mathrm{int}^{\mathrm{el}} &= \int\dd{\vb*{r}}\ev{\hat{\rho}_{\QM}(\vb*{r})\phi_{\MM}(\vb*{r})}\\
		% E_\mathrm{int}^{\mathrm{rep}}&= \ev*{\hat{V}_\mathrm{rep}}
		E_\mathrm{int}^{} &= \int\dd{\vb*{r}}\ev{\hat{\rho}_{\QM}(\vb*{r})\phi_{\MM}(\vb*{r})} + \ev*{\hat{V}_\mathrm{rep}}
	\end{align}
	so they can both be treated as corrections to the one-electron term in the many-body Hamiltonian for the QM system. We will shortly see that this is no longer true when the polarizability of the MM environment is introduced in the QM/MM model. The aim of the DRF approach is to avoid such non-linearity and to fold the polarizable response directly into a correction to the many-body Hamiltonian.\cite{thole_quantum_1980,thole_direct_1982}
	\vspace{-10pt}
	\subsection{The classical polarization energy}
	\vspace{-10pt}
	The DRF Hamiltonian can be obtained by considering the energy of a classical system of polarizable sites, each with an induced dipole $\vb*{\mu}_A$, coupled to some external charge distribution $\rho(\vb*{r})$. The classical energy functional for this system $\mathcal{F}_{\mathrm{pol}}\equiv\mathcal{F}_{\mathrm{pol}}[\rho,\vb*{\mu}]$ is given by\cite{bondanza_polarizable_2020}
	\begin{align}
		\begin{split}
			\mathcal{F}_{\mathrm{pol}} = &-\sum_{A=1}^{N_{\mathrm{MM}}}\vb*{\mathcal{E}}_\rho(\vb*{R}_A^{\MM})\cdot \vb*{\mu}_A + \frac{1}{2}\sum_{A=1}^{N_{\mathrm{MM}}} \frac{1}{\alpha_A} \vb*{\mu}_A\cdot \vb*{\mu}_A \\
			&+\frac{1}{2}\sum_{A=1}^{N_{\mathrm{MM}}}\sum_{B=1}^{N_{\mathrm{MM}}}\vb*{\mu}_A^\transp \vb{T}_{AB} \vb*{\mu}_B
		\end{split}\\
		&= -\vb*{\mathcal{E}}[\rho] \vb*{\mu}  + \frac{1}{2}\vb*{\mu}(\vb*{\alpha}^{-1} + \vb{T})\vb*{\mu}.
	\end{align}
	where the electric field at $\vb*{R}$ is given by
	\begin{align}
		\vb*{\mathcal{E}}_\rho(\vb*{R}) = -\int \dd{\vb*{r}} \left(\nabla_{\vb*{r}}\frac{1}{\|\vb*{r}-\vb*{R}\|}\right)\rho(\vb*{r})
	\end{align}
	$\vb*{\alpha}$ is a diagonal matrix of atomic polarizabilities, and $\vb{T}_{AB}$ is the interaction kernel between dipoles $A$ and $B$, with $\vb{T}_{AA} = \vb{0}$,\cite{nicoli_assessing_2022} 
	\begin{align}
		\vb{T}_{AB} = -\frac{1}{R_{AB}^3}\left(\frac{3{\vb*{R}}_{AB}{\vb*{R}}_{AB}^\transp}{R_{AB}^2}-\vb*{1}\right) 
	\end{align}
	where $\vb*{R}_{AB} = \vb*{R}_{A}^{\MM} - \vb*{R}_{B}^{\MM}$ and $R_{AB} = \|\vb*{R}_{AB}\|$. The interaction kernel may also include Thole damping for dipoles in close proximity.\cite{nicoli_assessing_2022} The classical energy of this system is found by minimizing $\mathcal{F}$ with respect to each dipole moment component $\mu_{A\alpha}$. This yields the following solution for the induced dipole moments $\vb*{\mu}_\mathrm{min}$ that minimize the polarization energy in response to the electric fields $\vb*{\mathcal{E}}$
	\begin{align}\label{eq-ind-eq}
		\vb*{\mu}_{\mathrm{min}}[\rho] = (\vb*{\alpha} + \vb{T})^{-1} \vb*{\mathcal{E}}[\rho],
	\end{align}
	and substituting this into the equation for the total polarization energy we find 
	\begin{align}
		E_{\mathrm{pol}}[\rho] &= \min_{\vb*{\mu}} \mathcal{F}_\mathrm{pol}[\rho,\vb*{\mu}] \\
		&= -\frac{1}{2} \vb*{\mathcal{E}}[\rho]^\transp\vb{K}_\mathrm{D}\vb*{\mathcal{E}}[\rho].\label{eq-classical-pol-energy}
	\end{align}
	where $\vb{K}_\mathrm{D} = (\vb*{\alpha} + \vb{T})^{-1} $. We note that the electric field components at each of the polarizable sites $\mathcal{E}[\rho]$ are linear functionals of $\rho(\vb*{r})$, so overall the polarization energy is a quadratic functional of the charge density.
	
	An alternative treatment of the polarizable MM environment that we will consider is the fluctuating charge approach.\cite{rick_dynamical_1994} In this approach the polarizability of the MM system is accounted for by allowing the charges on MM atoms to respond to external electrostatic potentials. In this formalism the polarization is described by a set of fluctuating charges $q_A$ rather than induced dipole moments $\vb*{\mu}_{A}$. The set-up is very similar to the dipole case, where we start by defining an energy functional $\mathcal{F}_{\mathrm{FQ}} \equiv \mathcal{F}_{\mathrm{FQ}}[\rho,\vb*{q}]$,
	\begin{align}
		\begin{split}
			\mathcal{F}_{\mathrm{FQ}} = &\sum_{A=1}^{N_\MM}(\phi_A[\rho]-\chi_A) q_A + \sum_{A=1}^{N_\MM} \frac{1}{2}\eta_A q_A^2 \\
			&+\frac{1}{2}\sum_{A=1}^{N_\MM}\sum_{B=1}^{N_\MM} q_A T_{AB}^{\mathrm{FQ}} q_B.
		\end{split}
	\end{align}
	where $\phi_A[\rho]$ is the electrostatic potential generated at site $A$ by the external charge density $\rho(\vb*{r})$,
	\begin{align}
		\phi_A[\rho] = \int\dd{\vb*{r}} \frac{1}{\|\vb*{r}-\vb*{R}^{\MM}_A\|} \rho(\vb*{r}),
	\end{align}
	$\chi_A$ is the electronegativity of atom $A$, $\eta_A$ is its chemical hardness and the interaction kernel $T_{AB}^{\mathrm{FQ}} = \eta_{AB}/\sqrt{1+\eta_{AB}R_{AB}^2}$ for $A\neq B$ and $T_{AA}^{\mathrm{FQ}} = 0$, where $\eta_AB = (\eta_A +\eta_B)/2$.
	
	Exactly as with the dipole polarization model, the final polarization energy is found by minimizing this energy functional with respect to the fluctuating charges $q_A$. The additional complication however is that the total charge of the full set of fluctuating charges (and also often certain subsets, such as individual molecules) is fixed at some value, and therefore the minimization has to be subject to this constraint.\cite{rick_dynamical_1994,lipparini_linear_2012} This slightly complicates the derivation of the final polarization energy which, nevertheless, is very similar to that of the induced dipole model,
	\begin{align}\label{eq-fq-energy}
		E_{\mathrm{FQ}}[\rho] = E_{\mathrm{FQ,0}} + \vb*{q}_0 \cdot \vb*{\phi}[\rho] - \frac{1}{2}\vb*{\phi}[\rho]^\transp\vb{K}_{\mathrm{FQ}}\vb*{\phi}[\rho],
	\end{align}
	where $E_\mathrm{FQ,0}$ is the energy of the system in the absence of external potentials, $\vb*{q}_0$ is the set of equilibrium charges in the absence of external potentials and $\vb*{K}_{\mathrm{FQ}}$ is an interaction kernel involving a matrix inverse similar to $\vb{K}_\mathrm{D}$. Because $\vb*{\phi}[\rho]$ is a linear functional of $\rho(\vb*{r})$, the fluctuating charge model polarization energy is a quadratic functional of the charge density, just like the induced dipole polarization model. In what follows for the sake of simplicity we will only focus on the induced dipole polarization model. Because the fluctuating charge polarization energy is functionally identical to the induced dipole one, the mean-field and DRF approaches described below can be extended trivially to the fluctuating charge model. 
	\vspace{-10pt}
	\subsection{Mean-field polarization}
	\vspace{-10pt}
	One way to account for the polarizability of the MM environment in a QM/MM model is to start with the classical polarization energy, Eq. (\ref{eq-classical-pol-energy}), and substitute the charge density $\rho(\vb*{r})$ with the expectation value of the charge density in the whole system,\cite{bondanza_polarizable_2020} i.e.
	\begin{align}
		\rho(\vb*{r}) \to \ev{\rho_\QM(\vb*{r})} + \rho_{\MM}(\vb*{r}).
	\end{align}
	With this substitution the QM/MM energy is minimized with respect to variations of the QM system wave function by solving the equations for the induced dipoles and electronic wave-function self-consistently,\cite{bondanza_polarizable_2020}
	\begin{subequations}\label{mf-pol-eqs}
		\begin{align}
			% \epsilon \ket{\Psi} &= (\hat{H}_{\QM} + \hat{V}_{\mathrm{int}}(\vb*{\mu}))\ket{\Psi} \\
			\lambda \Psi &=  \frac{\delta E_\mathrm{tot}[\Psi,\vb*{\mu}]}{\delta \Psi^*}\\
			\vb*{\mu} &= (\vb*{\alpha}^{-1}+\vb{T})^{-1}(\vb*{\mathcal{E}}[\ev{\rho_{\QM}}] + \vb*{\mathcal{E}}[{\rho_{\MM}}]),
		\end{align}
	\end{subequations}
	where $\lambda$ is a Lagrange multiplier ensuring $\braket{\Psi} = 1$. This set of equations defines the mean-field polarization method, which is the most common approach to accounting for polarization effects in QM/MM calculations. Although this approach is conceptually appealing in its simplicity, the mean-field energy functional is quadratic in the QM system observables, which means the energy does not arise from a simple eigenvalue equation. In addition to the practical difficulties of needing to self-consistently solve the Eq.~\eqref{mf-pol-eqs}, this also complicates the treatment of excited states.\cite{liu_conical_2022} For excited states one has a choice between fixing the polarization of the MM system in equilibrium with the QM ground state $\vb*{\mu} \to \vb*{\mu}[\ev{\rho_{\mathrm{gs}}}]$, which fails to account for the polarization response of the environment to the change in the QM density in the excited state, or one can self-consistently solve the equations again for the excited state density $\ev{\rho_{\mathrm{es}}(\vb*{r})}$. The latter choice however leads to non-orthogonal ground and excited state wave functions, which is clearly undesirable. As discussed by other authors, this ambiguity in how to treat excited states in the mean-field formulation can lead to severe problems in describing conical intersections in polarizable environments.\cite{liu_conical_2022} This motivates finding an alternative way to include polarization into QM/MM calculations. \hlt{Some alternatives that have been suggested include using linear response theory\cite{loco_qmmm_2016} and corrected linear response theory\cite{giovannini_electronic_2019}, as well as state averaged approaches\cite{hagras_polarizable_2018,song_state_2023}, although we will see that the DRF approach below has some important advantages over these approaches.}
	\vspace{-20pt}
	\subsection{Direct reaction field polarization}
	\vspace{-10pt}
	The DRF approach is an alternative to the mean-field treatment of the MM polarization.\cite{thole_quantum_1980,thole_direct_1982,liu_conical_2022,humeniuk_multistate_2024} In this approach, we take the expression for the classical polarization energy $E_\mathrm{pol}[\rho]$ and quantize it, by replacing the charge density of the QM system $\rho_{\QM}(\vb*{r})$ with the quantum mechanical charge density operator, $\hat{\rho}_\QM(\vb*{r})$,
	\begin{align}
		\hat{\rho}_\QM(\vb*{r}) = -\sum_{i=1}^{N_\mathrm{e}} \delta(\vb*{r} - \hat{\vb*{r}}_i) + \sum_{A=1}^{N_{\QM}} Z_A \delta(\vb*{r}-\vb*{R}_A) .
	\end{align}
	Accordingly, the polarization energy is treated as an additional term in the QM Hamiltonian
	\begin{align}
		\hat{V}_\mathrm{pol} = -\frac{1}{2} (\hat{\vb*{\mathcal{E}}}_\mathrm{QM} + \vb*{\mathcal{E}}_\mathrm{MM})^\transp\vb{K}_\mathrm{D}(\hat{\vb*{\mathcal{E}}}_\mathrm{QM} + \vb*{\mathcal{E}}_\mathrm{MM})
	\end{align}
	where $\hat{\vb*{\mathcal{E}}}_\QM = \vb*{\mathcal{E}}[\hat{\rho}_\mathrm{QM}]$ and ${\vb*{\mathcal{E}}}_\MM = \vb*{\mathcal{E}}[{\rho}_\mathrm{MM}]$. We note that this introduces not only new one-electron terms into the QM Hamiltonian, but also new two-electron terms
	\begin{align}
		\hat{V}_\mathrm{pol}^{(2)} = -\frac{1}{2}\sum_{i\neq j} \hat{\vb*{\mathcal{E}}}_{\QM,i}^\transp \vb{K}_\mathrm{D} \hat{\vb*{\mathcal{E}}}_{\QM,j}
	\end{align}
	where $\hat{\vb*{\mathcal{E}}}_{\QM,i} = \vb*{\mathcal{E}}[\hat{\rho}_{\mathrm{e},i}]$ and $\hat{\rho}_{\mathrm{e},i} = -\delta(\vb*{r}-\hat{\vb*{r}}_i)$. Because the MM polarization energy is included at the level of the many-body Hamiltonian for the QM system, ground and excited states can be treated on equal footing, avoiding issues of non-orthogonality. State specific polarization is automatically folded in with the DRF approach by introducing the MM polarization at the level of the Hamiltonian. Furthermore the DRF approach accounts (approximately) for dispersion interactions between the QM system and the MM system.\cite{liu_conical_2022,humeniuk_multistate_2024} This can be understood by considering that the expectation value of the DRF polarization energy involves not only the expectation value of the electric fields generated by the QM system but also their fluctuations. 
	%$\ev*{\hat{\vb*{\mathcal{E}}}_\QM}$ but also the fluctuations in the QM system electric fields $\ev*{\delta\hat{\vb*{\mathcal{E}}}_\QM^2} = \ev*{(\hat{\vb*{\mathcal{E}}}_\QM - \ev*{\hat{\vb*{\mathcal{E}}}_\QM})^2}$. 
	
	One shortcoming of the DRF approach is that evaluating matrix elements of $\hat{V}_\mathrm{pol}$ involves the full matrix inverse $\vb{K}_\mathrm{D} = (\vb*{\alpha} + \vb{T})^{-1}$. Evaluation and storage of this matrix can become limiting when the number of polarizable degrees of freedom becomes large. In general construction of this matrix will scale as $\mathcal{O}(N_{\MM}^3)$ and it requires $\mathcal{O}(N_{\MM}^2)$ storage. Below we suggest a simple and physically motivated approximation to the DRF Hamiltonian which can significantly improve its efficiency, particularly for large QM/MM systems. 
	\vspace{-10pt}
	\subsection{ESPF multipole operators}
	\vspace{-10pt}
	The starting point for deriving an efficient approximation to the full DRF Hamiltonian is to consider expanding the QM charge density operator with an atom-centred multipole expansion,\cite{ferre_approximate_2002}
	\begin{align}
		\begin{split}
			\hat{\rho}_{\QM}(\vb*{r}) \approx \sum_{A=1}^{N_\QM} &\bigg((\hat{Q}^A+Z_A) \delta(\vb*{r}-\vb*{R}_A) \\
			&- \hat{\vb*{\mu}}^A\cdot\nabla_{\vb*{r}}\delta(\vb*{r}-\vb*{R}_A) + \cdots\bigg)
		\end{split}
	\end{align}
	where $\hat{Q}^A$ is the atom centred electronic charge operator for QM atom $A$ and $\hat{\vb*{\mu}}^A$ is the atom centred electronic dipole moment operator. The expansion can in principle be continued to quadrupole and higher moments but here we only consider contributions up to atom-centred dipoles. In order to simplify the equations below, we introduce the following compact notation for the atom-centred multipole expansion,
	\begin{align}
		\hat{\rho}_{\QM}(\vb*{r}) &\approx \sum_{a=1}^{N_\mathcal{Q}}\hat{\mathcal{Q}}_a\rho_a(\vb*{r}) + \sum_{A=1}^{N_\QM} Z_A \delta(\vb*{r}-\vb*{R}_A) \\
		&= \sum_{a=1}^{N_\mathcal{Q}}(\hat{\mathcal{Q}}_a+\mathcal{Z}_a)\rho_a(\vb*{r})
	\end{align}
	where the index $a$ runs over all QM sites and multipole expansion components (charges and dipole moment components). We use $\mathcal{Z}_a$ to denote the nuclear charge contribution to each atom-centred multipole $a$, such that $\mathcal{Z}_a = Z_A$ when $a$ corresponds to an atom centred charge and $\mathcal{Z}_a = 0$ otherwise. For example if only charges are included in the expansion then $a$ runs from 1 to $N_\mathcal{Q} = N_\QM$, $\hat{\mathcal{Q}}_A = \hat{Q}^A$ and $\rho_A(\vb*{r}) = \delta(\vb*{r} - \vb*{R}_A)$, and when dipoles are included $a$ runs from 1 to $N_\mathcal{Q} = 4 N_\QM$.
	
	% \textcolor{blue}{NF: I have the feeling the above change of notation is not clear enough. For instance, when dipoles are considered, is $a$ running from 1 to $4\times N_{QM}$?}
	
	The charge operators $\hat{Q}^A$ and dipole operators $\hat{\vb*{\mu}}^A$ can be decomposed into a sum of one electron terms, $\hat{Q}^A_i$ and $\hat{\vb*{\mu}}^A_i$. There are many ways of defining these multipole operators, but since we are interested in using these operators to describe electrostatic interactions, we use an electrostatic potential fitting procedure to derive these operators. The matrix elements of the operators $Q^A_{\mu\nu}$ and $\mu_{\alpha,\mu\nu}^A$ in a given basis set are found by minimising difference between the electrostatic potential (ESP) generated by the set of charges and dipoles and the exact ESP matrix elements at a set of $N_\mathrm{grid}$ grid points $\vb*{r}_k$\cite{ferre_approximate_2002,huix-rotllant_analytic_2021}
	\begin{align}
		\mathcal{L}_{\mu\nu} = \sum_{k=1}^{N_\mathrm{grid}} \bigg( \sum_{a=1}^{N_\mathcal{Q}}\phi_a(\vb*{r}_k) \tilde{\mathcal{Q}}_{a,\mu\nu} - \Phi_{\mu\nu}(\vb*{r}_k)\bigg)^2 \label{eq-ESPF}
	\end{align}
	where the ESPF matrix elements are
	\begin{align}
		\Phi_{\mu\nu}(\vb*{r}_k) = \frac{1}{2}\int\dd{\vb*{r}}\chi_\mu(\vb*{r}) \frac{1}{\|\vb*{r} - \vb*{r}_k\|}\chi_{\nu}(\vb*{r})
	\end{align}
	and $\phi_a(\vb*{r}_k)$ is the potential generated by a unit multipole $a$ at $\vb*{r}_k$,
	\begin{align}
		\phi_a(\vb*{r}_k) = \int\dd{\vb*{r}} \frac{1}{\|\vb*{r}-\vb*{r}_k\|} \rho_a(\vb*{r})
	\end{align}
	so for atom centred charges this is given by
	\begin{align}
		\phi_A(\vb*{r}_k) = \frac{1}{\|\vb*{r}_k - \vb*{R}_A\|}
	\end{align}
	and for dipoles this is given by
	\begin{align}
		\phi_{A,\alpha}(\vb*{r}_k) = \frac{[\vb*{r}_k - \vb*{R}_A]_\alpha}{\|\vb*{r}_k - \vb*{R}_A\|^3}.
	\end{align}
	In this work we use atom-centred Lebedev grids as the choice of grid points $\vb*{r}_k$.\cite{huix-rotllant_analytic_2021} The multipole components which minimise Eq. (\ref{eq-ESPF}) are given by
	\begin{align}
		\tilde{\vb*{\mathcal{Q}}}_{\mu\nu} = (\vb*{\phi}^\transp\vb*{\phi})^{-1} \vb*{\phi}^\transp\vb*{\Phi}_{\mu\nu}.
	\end{align}
	The fitted multipole operators should obey the exact sum rules Eq.~\eqref{eq-S-sum} and Eq.~\eqref{eq-mu-sum}, below to ensure they reproduce the electrostatic potential of the full charge distribution at long range\cite{huix-rotllant_analytic_2021}
	\begin{align}
		- S_{\mu\nu} &= \sum_{A=1}^{N_\QM} Q^A_{\mu\nu}\label{eq-S-sum}   \\
		- \vb*{r}_{\mu\nu} &= \sum_{A=1}^{N_\QM} (Q^A_{\mu\nu} \vb*{R}_A + \vb*{\mu}^A_{\mu\nu}) \label{eq-mu-sum}
	\end{align}
	where $S_{\mu\nu} = \braket*{\chi_\mu}{\chi_\nu}$ and $\vb*{r}_{\mu\nu} =\mel*{\chi_\mu}{\hat{\vb*{r}}}{\chi_\nu} $. Eq.~\eqref{eq-S-sum} ensures the charge operators reproduce the total charge of the molecule, and Eq.~\eqref{eq-mu-sum} ensures the total dipole moment is reproduced, thereby guaranteeing the fitted operators reproduce long-range electrostatic interactions correctly. 
	The fitted matrix elements are corrected as follows to enforce these conditions and give the final definitions of the atom-centred charge and dipole operators,\cite{huix-rotllant_analytic_2021}
	\begin{align}
		Q^A_{\mu\nu} &= \tilde{Q}^A_{\mu\nu} - \frac{1}{N_{\QM}}\sum_{B=1}^{N_\QM}(\tilde{Q}^B_{\mu\nu} +S_{\mu\nu}) \\
		\vb*{\mu}^A_{\mu\nu} &= \tilde{\vb*{\mu}}^A_{\mu\nu} - \frac{1}{N_{\QM}}\sum_{B=1}^{N_\QM}(\tilde{\vb*{\mu}}^B_{\mu\nu} + Q^B_{\mu\nu} \vb*{R}_B +\vb*{r}_{\mu\nu}).
	\end{align}
	\hlt{These corrections correspond to the minimal perturbations to the unconstrained multipole operators (in a least-squares sense) which enforce the constraints (a test of corrected vs. uncorrected ESPF operators is given in the supporting information).} The utility of approximating the charge density operator in this way will be expanded upon in the next section.
	\vspace{-10pt}
	\subsection{DRF with multipole operators}
	\vspace{-10pt}
	In order to derive an approximate but efficient formulation of the DRF Hamiltonian, we insert out multipole based approximation to QM region charge density into the the QM electric field operator (which define each component of the vector $\hat{\vb*{\mathcal{E}}}_\QM$)
	\begin{align}
		\hat{\vb*{\mathcal{E}}}_\QM(\vb*{R}) &= - \int \dd{\vb*{r}}\bigg(\nabla_{\vb*{r}}\frac{1}{\|\vb*{r}-\vb*{R}\|}\bigg)\hat{\rho}_\QM(\vb*{r}) \nonumber\\
		&\approx \sum_a \int \dd{\vb*{r}}\bigg(\!\!-\nabla_{\vb*{r}}\frac{1}{\|\vb*{r}-\vb*{R}\|}\bigg){\rho}_a(\vb*{r})(\hat{\mathcal{Q}}_a +\mathcal{Z}_a) \nonumber\\
		&= \sum_a (\hat{\mathcal{Q}}_a + \mathcal{Z}_a) \vb*{f}_a(\vb*{R})
	\end{align}
	where $\vb*{f}_a(\vb*{R})$ is the electric field generated by multipole component $a$ with unit magnitude, and using $\vb*{f}_a$ to denote the vertically stacked vector of all of these fields $\vb*{f}_a(\vb*{R}_A^\MM)$ for each MM polarizable site, the $\hat{\vb*{\mathcal{E}}}_\QM$ operator can be expressed as
	\begin{align}
		\hat{\vb*{\mathcal{E}}}_\QM \approx \sum_a (\hat{\mathcal{Q}}_a + \mathcal{Z}_a) \vb*{f}_a.
	\end{align}
	Substituting this into the DRF Hamiltonian we obtain
	\begin{align}
		\hat{V}_\mathrm{pol} &= V_{\mathrm{pol},0} + \hat{V}_\mathrm{pol}^{(1)} +  \hat{V}_\mathrm{pol}^{(2)} 
	\end{align}
	where the $V_{\mathrm{pol},0}$ corresponds to the polarization energy of the MM system in the absence of the QM system electrons,
	\begin{align}\label{eq-Vpol0}
		\begin{split}
			V_{\mathrm{pol},0} &= -\frac{1}{2} \Ef_\MM^\transp \vb{K}_\mathrm{D} \Ef_\MM \\
			&-\frac{1}{2}\sum_{ab}\mathcal{Z}_a U_{\QM}^{ab} \mathcal{Z}_b -\sum_{ab}\mathcal{Z}_a U_{\QM/\MM}^{a}.
		\end{split}
	\end{align}
	The 1-electron term $\hat{V}^{(1)}_\mathrm{pol}$ can be splitted into a term arising from the interaction of the MM dipoles induced by the MM electric fields interacting with the QM density, and a self-interaction term arising from the response of the induced dipoles to the charges in the QM system
	\begin{align}
		\hat{V}_\mathrm{pol}^{(1)} &= \hat{V}_\mathrm{ind}^{(1)} + \hat{V}_\mathrm{self}^{(1)}\\
		\hat{V}_\mathrm{ind}^{(1)} &= - \sum_a \bigg(U^a_{\QM/\MM}+\sum_b U_{\QM}^{ab}\mathcal{Z}_b\bigg) \hat{\mathcal{Q}}_a \\
		\hat{V}_\mathrm{self}^{(1)} &= -\frac{1}{2}\sum_{i=1}^{N_\mathrm{e}} \sum_{a,b}U_{\QM}^{ab}\hat{\mathcal{Q}}_{a,i}\hat{\mathcal{Q}}_{b,i}.
		% \big(\hat{\mathcal{Q}}_{a,i}+\frac{\mathcal{Z}_a}{N_\mathrm{e}}\big)\big(\hat{\mathcal{Q}}_{b,i}+\frac{Z_b}{N_\mathrm{e}}\big)
	\end{align}
	The 2-electron contribution arises from the interaction of one electron with the dipoles induced by another electron in the system,
	\begin{align}
		\hat{V}_\mathrm{pol}^{(2)} &= -\frac{1}{2}\sum_{i\neq j}^{N_\mathrm{e}} \sum_{a,b}U_{\QM}^{ab}\hat{\mathcal{Q}}_{a,i}\hat{\mathcal{Q}}_{b,j}
	\end{align}
	The variables $U^a_{\QM/\MM}$ and $U_{\QM}^{ab}$ are given by
	\begin{align}
		U_{\QM/\MM}^a &= \Ef_\MM^\transp \vb{K}_\mathrm{D} \vb*{f}_a \\
		U_{\QM}^{ab} &= \vb*{f}_a^\transp \vb{K}_\mathrm{D} \vb*{f}_b.
	\end{align}
	This means we do not have to evaluate the full matrix $\vb{K}_\mathrm{D} = (\vb*{\alpha}^{-1} + \vb{T})^{-1}$, but instead we only need to evaluate $\vb*{K}_\mathrm{D}\vb*{v}$, for all $\vb*{v} = \vb*{f}_a$ and $\vb*{v} =\Ef_\MM$, and then evaluate and store the inner products of these with $\Ef_\MM$ and $\vb*{f}_a$, together with constructing and storing the atom-centred multipole operators $\hat{\mathcal{Q}}_a$ in a given basis set. We note that $\vb*{K}_\mathrm{D}\vb*{f}_a$ is the solution to the dipole induction equations for a field generated by atom-centred multipole $a$, and therefore setting up the approximate DRF Hamiltonian only involves solving the induction equations for a small number of input multipoles at the QM sites. Using this procedure we therefore avoid constructing the full matrix inverse $(\vb*{\alpha}^{-1} + \vb{T})^{-1}$, and instead we only require the solution to the MM system induction equations [Eq.~\eqref{eq-ind-eq}] for $N_\mathcal{Q}+1$ different external charge densities. The induction equations can be solved iteratively, requiring matrix-vector evaluations that scale at most as $\mathcal{O}(N_\MM^2)$ per iteration if the interactions are not truncated at finite range. When long-range interactions are truncated or treated using Ewald sum type methods this scaling can be reduced further,\cite{toukmaji_ewald_1996,shan_gaussian_2005,darden_particle_1993,chollet_ankh_2023} and several methods exist for very quickly converging the solution to these equations.\cite{wang_fast_2005,nocito_reduced_2019,gubler_accelerating_2024} This enables ESPF-DRF to be applied in principle to very large polarizable MM systems. \hlt{Avoiding construction of the full $\vb{K}_\mathrm{D}$ by iteratively solving a set of linear equations to construct the IEDRF Hamiltonian has been suggested in Ref.~\onlinecite{humeniuk_multistate_2024} for IEDRF. Thus for IEDRF $\vb*{K}_\mathrm{D}\vb*{v}$ needs to be evaluated $N_{\mathrm{AO}}(N_{\mathrm{AO}}+1)/2$ times (where $N_\mathrm{AO}$ is the size of the atomic orbtial basis used int he calculation), compared to $N_\mathcal{Q}$ times for ESPF-DRF (where $N_\mathcal{Q} = n_\mathcal{Q} N_\MM $, where $n_\mathcal{Q}$ is the number of multipole components per atom.) Therefore ESPF-DRF offers a speed-up of roughly $N_{\mathrm{AO}}^2/2N_\mathcal{Q}$ in cases where solving the induction equations for the MM region is the limiting step. In Ref.~\onlinecite{humeniuk_multistate_2024} it was found that the final contraction in evaluation of the exchange matrix after the induction equations have been solved was often the rate-limiting step in calculations, which scales as $\mathcal{O}(3N_\MM N_{\mathrm{AO}}^3)$. As we show in the supporting information, the equivalent step is ESPF-DRF scales as $\mathcal{O}(N_\mathcal{Q} N_{\mathrm{AO}}^3)$, so ESPF-DRF offers a speed-up of $\sim 3N_\MM / N_\mathcal{Q} $ when the DRF exchange matrix evaluation is rate-limiting. More detailed scaling analysis of ESPF-DRF and IEDRF is given in the supporting information.} %In our implementation where no cut-off is introduced for dipole-dipole interactions, we have found empirically that the computation time scales approximately as $\mathcal{O}(\log(N_\MM)^2 N_\MM^2)$, whereas when using the full matrix inverse (which would have the same scaling as IEDRF) it is approximately $\mathcal{O}(N_\MM^{2.85})$. When no cut-off is applied to the dipole-dipole interactions the minimum theoretical scaling is $\mathcal{O}(N_\MM^2)$ but if cut-offs and better were used lower scaling could be achieved. 
	% Could include the above sentence^
	%[Also note that in a non-orthogonal basis the operator product $\hat{\mathcal{Q}}_{a,i}\hat{\mathcal{Q}}_{b,i}$ is represented by $\vb*{\mathcal{Q}}_a \vb{S}^{-1} \vb*{\mathcal{Q}}_b$, where $[\vb*{\mathcal{Q}}_a]_{\mu\nu} = \mathcal{Q}_{\mu\nu}$ is the matrix representation of $\hat{\mathcal{Q}}_{a}$ in the non-orthogonal basis and $[\vb{S}]_{\mu\nu} = S_{\mu\nu}$ is the overlap matrix.]
	
	Essentially the same idea can be applied with a fluctuating charge model for the MM system polarizability. We just replace the charge density $\rho$ with the sum of MM and approximate atom-centred multipole QM charge density operator in the expression for $E_\mathrm{FQ}[\rho]$. This also avoids construction of the full $\vb{K}_{\mathrm{FQ}}$ matrix which (as shown in the appendix) involves a large matrix inverse. The equations for the ESPF-DRF method with FQ polarization are detailed in Appendix \ref{app-fq-drf} together with a brief derivation. The main additional complexity arises from enforcing total charge constraints in the MM region, which we also address in this appendix.
	\vspace{-10pt}
	\subsection{Practical considerations and implementation}
	\vspace{-10pt}
	Before demonstrating the accuracy and utility of using ESPF charge operators with the DRF polarizability, we consider a few practical considerations in implementing the DRF method. As noted by other authors, the DRF Hamiltonian contains divergences unless damping is added. Whilst using an atom-centred multipole expansion automatically avoids the divergences, the interaction energies can still become too large when MM atoms comes very close to the QM region. In order to avoid these problems, we replace the full Coulomb interaction at MM atom $B$ with a damped version from a QM multipole at atom $A$ with
	\begin{align}\label{eq-damped-coul}
		\frac{1}{\|\vb*{r}-\vb*{R}_B^{\MM}\|} \to \frac{1}{\left(\|\vb*{r}-\vb*{R}_B^{\MM}\|^m + \left(R^{\mathrm{damp}}_{A,B}\right)^m\right)^{1/m}}
	\end{align}
	where the damping radius is chosen to be the sum of covalent radii of the two atoms $R_{A,B}^\mathrm{damp} = R_{A}^{\mathrm{cov}}+R_{B}^{\mathrm{cov}}$ and $m$ is set to $m=6$ \hlt{(similar to damped QM/MM interactions used in some other implementations\cite{laino_efficient_2005})}. The covalent radii are taken directly from the PySCF radii module, with values from Ref.~\onlinecite{cordero_covalent_2008}, unless otherwise stated. \hlt{For metal cations the covalent radii of the neutral atoms are generally too large, so in these cases we use the effective ionic radii instead.\cite{shannon_revised_1976}} This damping scheme is inspired by Becke-Johnson damping used in the empirical dispersion corrections to DFT methods,\cite{grimme_effect_2011} but it is obviously not the only possibility, for example Thole type damping could be used as an alternative.\cite{nicoli_assessing_2022} This only affects how exact the vector $\vb*{f}_a$ is constructed. \hlt{The sensitivity of the ESPF-DRF method to the choice of damping parameter is explored briefly in the supporting information.}
	
	Another practical issue worth considering is how to construct the modified two-electron integrals and the resulting modifications to the Coulomb and exchange matrices from the modified electron-electron interaction. Firstly we note that the modified two-electron integrals are given by
	\begin{align}
		[\mu\mu'|\nu\nu']_\mathrm{pol} %&= \frac{1}{2}\mel{\chi_\mu(1)\chi_{\nu}(2)}{\hat{V}^{(2)}_{\mathrm{pol}}}{\chi_{\mu'}(1)\chi_{\nu'}(2)} \\
		&= -\frac{1}{2}\sum_{a,b}U^{ab}_{\QM} \mathcal{Q}_{a,\mu\mu'}\mathcal{Q}_{b,\nu\nu'}.
	\end{align}
	With the atom-centred dipole approximation we can directly avoid constructing the full four-index tensor of the two-electron integrals when constructing the Coulomb and exchange matrices (as is generally possible for DRF methods due to its tensor-factorizable form as noted in Ref.~\onlinecite{humeniuk_multistate_2024}). For a given input one-particle reduced density matrix $\vb{D}$, the modified Coulomb and exchange matrices are given by 
	\begin{align}
		[\vb{J}_\mathrm{pol}(\vb{D})]_{\mu\nu} &= -\sum_a \mathcal{Q}_{a,\mu\nu}\sum_{b,\mu'\nu'} U_{\QM}^{ab} D_{\nu'\mu'}\mathcal{Q}_{b,\nu'\mu'}\\
		[\vb{K}_\mathrm{pol}(\vb{D})]_{\mu\nu} &= -\sum_{a,\nu'} \mathcal{Q}_{a,\mu\nu'}\sum_{b,\mu'} U_{\QM}^{ab} D_{\nu'\mu'}\mathcal{Q}_{b,\nu\mu'}.
	\end{align}
	These are also used in the calculation of the response functions needed in CIS/TD(A)-DFT methods. Our ESPF-DRF procedure avoids the need for special two-electron integrals as is needed in IEDRF\cite{liu_conical_2022,humeniuk_efficient_2022,humeniuk_multistate_2024} and the modified integrals are straighforward to implement in existing electronic structure codes.
	
	Combining DRF with Kohn-Sham DFT requires some brief consideration. Formally the DRF method alters the electron-electron interaction and should therefore change the exchange-correlation density functional. For simplicity, as done previously by others,\cite{humeniuk_multistate_2024} in order to combine DRF and KS-DFT we simply add the DRF Hartree-Fock/exact-exchange polarization energy to the DFT energy, 
	\begin{align}
		\begin{split}
			E_\mathrm{tot} &= E_\mathrm{QM,DFT} + E_\MM \\
			&\ \ \ + E_\mathrm{int}^\mathrm{el} + E_\mathrm{int}^\mathrm{rep} + E_\mathrm{DRF,HF} 
		\end{split}\\
		E_\mathrm{DRF,HF} &= \ev{\hat{V}_\mathrm{pol}}{\Phi_\mathrm{KS}}
	\end{align}
	where $\ket{\Phi_\mathrm{KS}}$ is the Slater determinant of Kohn-Sham orbitals. This ignores any potential effect of the altered electron-electron interaction on exchange/correlation functional beyond this simple additive model, but it is a practical approximation which is easily combined with existing electronic structure software.
	
	\hlt{Although the ESPF charge operator machinery is not wide-spread, it only requires standard one-electron integrals and ESPF operators can be obtained with a relatively simple modification of codes for calculating ESP charges (e.g. RESP or CHELPG charges) which are routinely used in QM/MM codes.} We have implemented ESPF-DRF as a Python module \texttt{PyESPF}\cite{pyespf} modifying the PySCF\cite{sun_pyscf_2018,sun_libcint_2015,sun_recent_2020} routines for all electronic structure calculations, covering the combinations of  with Hartree-Fock(HF)/density functional theory (DFT) as well as configuration interaction singles (CIS)/time-dependent DFT (TDDFT) for excited states. \texttt{PyESPF} is freely available at \url{www.github.com/tomfay/PyESPF}.
	
	\section{Results}
	
	\subsection{Small bimolecular systems}
	
	In order to evaluate the strengths and potential shortcomings of the ESPF-DRF method, we have considered the interaction energies of small molecular systems of increasing complexity. These systems provide a way to test the multipole expansion approximation, where we test the ESPF-DRF method with the expansion truncated at atom-centred charges [denoted DRF(Q)] and with the expansion truncated at dipoles [denoted DRF(Q+$\muup$)]. %In each case we have performed high level quantum chemical calculations to obtain reference interaction energies, primarily the CCSD(T)-F12 method with the cc-pVQZ-F12 basis set. Use of the explicitly correlated F12 method reduces the potential effect of basis-set superposition error in the intermolecular interaction energy calculations. For excited states we have used the EOM-CCSD and CIS(D) methods, which can account for differences in long-range dispersion interactions in ground and excited electronic states. 
	\hltt{Since we are ultimately concerned with the accuracy of our ESPF-DRF methodology compared to exact calculations, we focus here primarily on direct comparisons between high-level electronic structure methods and ESPF-DRF, although we also provide a comparison between ESPF-DRF and available IEDRF data below as well. } All reference calculations were performed with Orca 6.\cite{neese_orca_2012,neese_software_2022,neese_span_2023} %In each case we use the PBE0 functional to account for exchange and correlation effects in the QM system, although we find the interaction energies only have a small dependence on the choice of method (Hartree-Fock or density functional approximation) used to describe the QM region.
	
	\begin{figure}[t]
		\centering
		\includegraphics[width=0.9\linewidth]{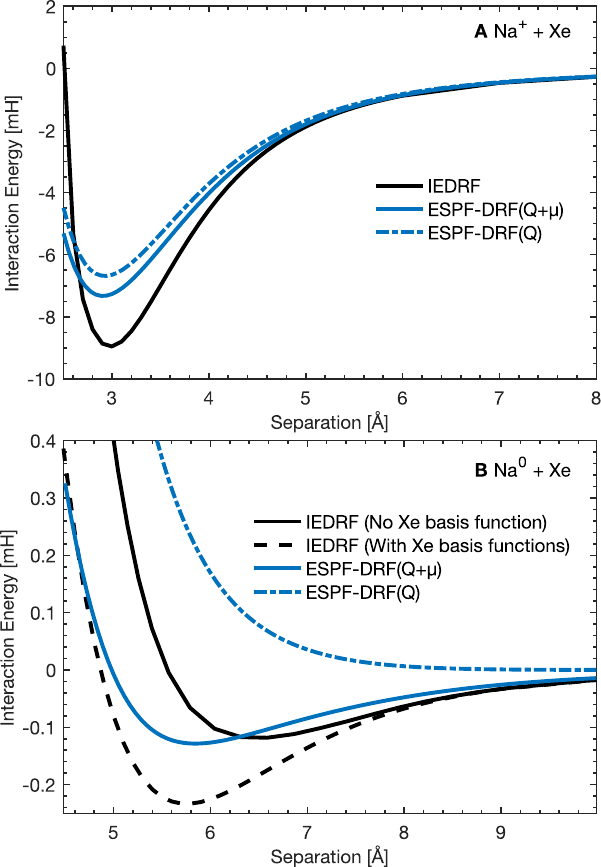}
		\caption{\hltt{ A) $\ce{Na+ + Xe}$  and B) $\ce{Na^0 + Xe}$ potential energy curves calculated with ESPF-DRF methods and with IEDRF (with data taken from Ref.~\onlinecite{liu_conical_2022}). In B) the black solid line corresponds to the IEDRF calculation with basis aug-cc-pV5Z-PP functions on Xe (with $f$ and higher angular momentum functions removed), and the dashed line corresponds to a calculation without Xe basis function. The ESPF-DRF calculations only include basis functions on Na in both cases. } }
		\label{fig-Na-Xe}
	\end{figure}
	
	\hltt{\subsubsection{Comparison of ESPF-DRF and IEDRF}
	
	Before evaluating the accuracy of ESPF-DRF against high-level reference calculations, it is worth comparing ESPF-DRF and its parent IEDRF method. In Fig.~\ref{fig-Na-Xe} we compare the potential energy curves for \ce{Na^+ + Xe} (Fig.~\ref{fig-Na-Xe}A) and \ce{Na^0 + Xe} (Fig.~\ref{fig-Na-Xe}B) calculated with ESPF-DRF to the IEDRF potential energy curves from Ref.~\onlinecite{liu_conical_2022}. In all calculations the QM region is treated with HF and the aug-cc-pV5Z basis with $f$ and higher orbital angular momentum functions removed. For \ce{Na^0 + Xe} the IEDRF calculations in Ref.~\onlinecite{liu_conical_2022} were run with and without aug-cc-pV5Z-PP basis functions on Xe (again with $f$ and higher angular momentum functions removed), whilst all ESPF-DRF calculations only include basis functions on Na. In all calculations the same damping function for electrostatic interactions and pseudo-potential treatment of Pauli repulsion were used. 
	
	Comparing first the ESPF-DRF [DRF(Q+$\upmu$)] with the corresponding IEDRF calculations in Fig.~\ref{fig-Na-Xe} we see, as expected, that at long range all calculations agree. At intermediate range, we see that the DRF(Q+$\upmu$) calculations underestimate the attraction between the QM \ce{Na} cation/atom and the MM Xe atom, which we attribute to the neglected higher-order multipoles in the QM region which become more important at shorter range. At very short range however DRF(Q+$\upmu$) overestimates the attractive polarization energy between the QM and MM regions. This is because the QM region charge density is treated as a point multipole, but at such short range the finite size of the QM region becomes important, introducing additional damping to the interaction which is naturally included in the IEDRF approach. This is why in all other calculations we describe the QM-MM interactions with the damped Coulomb interaction in Eq.~\ref{eq-damped-coul} which accounts for the finite size of both the QM and MM atoms. We see also that the DRF(Q) method is almost as accurate as DRF(Q+$\upmu$) for the \ce{Na^+ + Xe} case, but for \ce{Na^0 + Xe} it is purely repulsive. This is because the DRF(Q) method cannot capture the dispersion interaction between a single QM site and the MM region. The poorer performance of the DRF(Q+$\upmu$) for the neutral \ce{Na^0 + Xe} case can likely be attributed to the highly diffuse and polarizable  3$s$ orbital occupied in \ce{Na^0}, as is evidenced by the large difference between IEDRF calculations with and without basis functions on Xe.
	
	}
	
	\begin{figure*}
		\centering
		\includegraphics[width=0.9\textwidth]{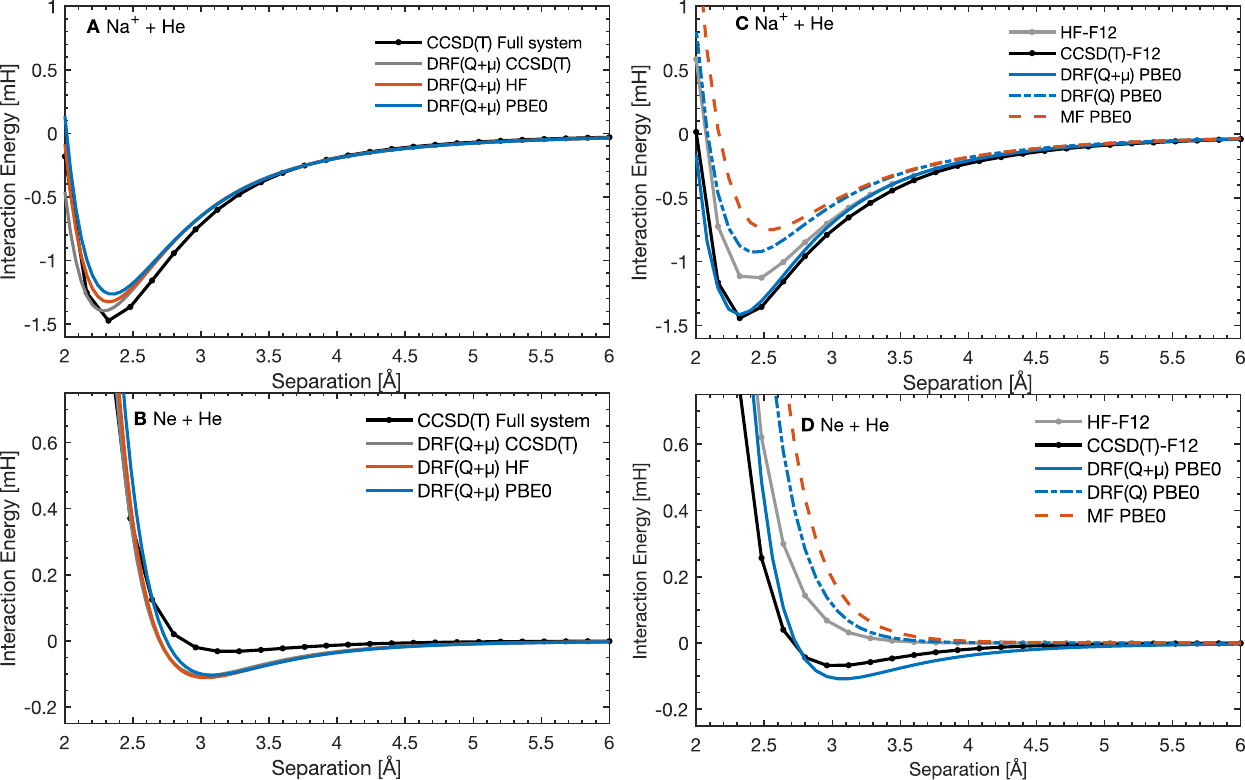}
		\caption{A) \ce{Na+ + He} and B) \ce{Ne + He} interaction energy curves calculated at the CCSD(T)/aug-ano-pVQZ level of theory for the full system (black line) or using ESPF-DRF with the QM system treated with CCSD(T) (grey line) HF (red line) and PBE0 (blue line). C) $\ce{Na+ + He}$ potential energy curves calculated with HF-F12/cc-pVQZ-F12 (grey solid line), CCSD(T)-F12/cc-pVQZ-F12 (black solid like), ESPF-DRF with charge and dipole operators (DRF(Q+$\muup$): blue solid line) and just charge operators (blue dot-dash line) and ESPF mean-field theory (DRF(Q): red dashed line) all using using PBE0/cc-pVQZ-F12. D) Same for the $\ce{Ne + He}$ system. \hlt{Note in this case the covalent radius for $\ce{Na+}$ is set to its ionic radius of 1.02 \AA\ for the damping to better reflect the small size of the $\ce{Na+}$ ion. All other covalent radii are fixed at their values in Ref.~\onlinecite{cordero_covalent_2008}.} }
		\label{fig-A-He}
	\end{figure*}
	
%	\hltt{
%	\subsubsection{Atomic interaction energies}
%	
%	Having examined the ESPF-DRF approach above, compared to reference IEDRF calculations, we now turn to our main objective: comparing ESPF-DRF to high-level full system reference calculations. To this end we have calculated CCSD(T)/aug-cc-pVQZ interaction energies for \ce{Li^+ + He} and \ce{He + He}, where in both cases one He atom is treated as the MM system, whilst the remaining atom, either \ce{Li^+} or \ce{He} are treated as the QM system. The QM system is treated with the same aug-cc-pVQZ basis set as in the full system reference calculations. The MM He atom is assigned a polarizability of $\alpha_{\ce{He}} = 1.38411\ \mathrm{a}_0^3 $, the polarizability of the isolated atom at the CCSD(T)/aug-cc-pVQZ level of theory. 	
%	
%	The results comparing the reference and ESPF-DRF calculations, where the QM system is treated at the CCSD(T) level of theory, are shown in Fig.~\ref{fig-lihe-he}A and B for \ce{Li^+ + He} and \ce{He + He} respectively. 
%	}
	\subsubsection{Atomic interaction energies}
	
%	\begin{figure}[h]
%		\centering
%		\includegraphics[width=1\linewidth]{fig1-A-He-new.pdf}
%		\caption{ A) $\ce{Na+ + He}$ potential energy curves calculated with HF-F12/cc-pVQZ-F12 (grey solid line), CCSD(T)-F12/cc-pVQZ-F12 (black solid like), ESPF-DRF with charge and dipole operators (DRF(Q+$\muup$): blue solid line) and just charge operators (blue dot-dash line) and ESPF mean-field theory (DRF(Q): red dashed line) all using using PBE0/cc-pVQZ-F12. B) Same for the $\ce{Ne + He}$ system. \hlt{Note in this case the covalent radius for $\ce{Na+}$ is set to its ionic radius of 1.02 \AA for the damping to better reflect the small size of the $\ce{Na+}$ ion. All other covalent radii are fixed at their values in Ref.~\onlinecite{cordero_covalent_2008}.} }
%		\label{fig-A-He}
%	\end{figure}
	
	\hltt{First to evaluate the quality of the ESPF-DRF method, we have calculated the interaction energies of the isoelectronic \ce{Na+ + He} and \ce{Ne + He} systems, with reference calculations on the full system performed at the CCSD(T)/aug-ano-pVQZ\cite{neese_revisiting_2011} level of theory. The He atom is treated as a polarizable MM atom, with the polarizability of an isolated He atom at the CCSD(T)/aug-ano-pVQZ level of theory, $1.12323\ \text{a}_0^3$. The exact and ESPF-DRF(Q+$\upmu$) calculations are shown in Fig.~\ref{fig-A-He}A and B as the black and grey lines. In both cases we see good agreement between the full system and ESPF-DRF calculations for \ce{Na+ + He}, with slightly worse performance for \ce{Ne + He}. The worse performance for \ce{Ne + He} is expected, because it is known that the DRF approach overestimates the strength fo dispersion interactions,\cite{angyan_are_1990,liu_conical_2022} in this case manifesting as an overestimation of the binding energy by a factor of  approximately $3$. For these two examples we have also examined the role of the level of theory with which the QM system is treated in the ESPF-DRF calculations, by also calculating the interaction energies with HF and PBE0 methods. In both cases there is relatively little variation in the interaction energy with the treatment of electron correlation in the QM region apart from at short range, which suggests that the electron correlation more strongly effects electron repulsion, rather than longer range DRF polarization energies.}
	
%	First we consider the isoelectronic \ce{Na+ + He} and \ce{Ne + He} systems,
	 \hltt{Now turning to our main objective of comparing the ESPF-DRF approach to high-level reference calculations, we analyse the performance of ESPF-DRF against CCSD(T)-F12/cc-pVQZ-F12 reference calculations for the same two systems. Note that we find the \ce{Na+ + He} binding energy to be very similar but for \ce{Ne + He} the binding energy is noticeably larger for CCSD(T)-F12/cc-pVQZ-F12 compared to CCSD(T)/aug-ano-pVQZ, likely due to reduced repulsion for the CCSD(T)-F12 at short range arising from the F12 treatment of electron correlation.}
	The He atom is now treated at the MM level of theory, with the atomic polarizability of He set to \qty{1.20409}{a_0^3}, corresponding the the polarizability calculated at the CCSD(T)-F12/cc-pVQZ-F12 level of theory (the same as the reference interaction energies).\cite{adler_simple_2007,pavosevic_geminal-spanning_2014,hill_correlation_2010,peterson_systematically_2008} In Fig.~\ref{fig-A-He} we show the reference CCSD(T)-F12 and HF-F12 reference energies together with the ESPF-DRF and ESPF-MF QM/MM interaction energies calculated at the PBE0/cc-pVQZ-F12 level of theory.\cite{perdew_rationale_1996}
	
	Turning our attention first to the \ce{Na+ + He} in Fig.~\ref{fig-A-He}C, we see that all the QM/MM methods accurately capture the asymptotic behaviour of the interaction curve where $E_\mathrm{int}(R) = -\alpha_{\ce{He}}/2R^4$. The charge and dipole version of the ESPF-DRF [DRF(Q+$\muup$)] method predicts stronger binding at shorter range because it can additionally capture dispersion interactions between the \ce{Na+} and \ce{He} atom, which agrees qualitatively with the fact that the Hartree-Fock-F12 interaction curve (which neglects electron correlation and therefore dispersion interactions) has a lower binding energy than the CCSD(T)-F12 interaction curve. Conversely, the charge-based ESPF-DRF [DRF(Q)] cannot capture dispersion interactions in this case because it is not flexible enough to reproduce the interaction between \ce{He} and the dipole moment of the QM region. The underestimation of the binding energy with the QM/MM methods in this system in likely a result of the inadequacies of the Pauli repulsion model we have used, which cannot account for how the MM region electron density is distorted by Pauli repulsion, as is evidenced by disagreement between the ESPF-MF and HF-F12 interaction curves, \hlt{where we see that the ESPF-MF interaction energy becoems positive faster than the HF-F12 curve}. Furthermore at very small separations, in the presence of strongly varying electric potentials close to the \ce{Na+} ion, treating the MM electron density as a point dipole may be insufficient for describing the interaction energy. 
	
	Now let us consider the \ce{Ne + He} system, with results shown in Fig.~\ref{fig-A-He}D. It is notable in this system that the HF-F12 interaction energy is purely repulsive, because the interaction energy arises entirely from dispersion interactions which can only be captured with correlated electronic structure methods. We see that the charge ESPF-DRF and the ESPF-MF methods fail to capture the potential energy well and predict repulsive interaction curves, qualitatively similar to the HF-F12 curve. In the case of charge ESPF-DRF method [DRF(Q)], this is because the single atom-centred charge operator cannot capture dipole fluctuations of the QM system that give rise to the attractive dispersion interaction between \ce{Ne} and \ce{He}. Including the atom-centred dipole contribution in the ESPF-DRF method [DRF(Q+$\muup$)] allows the method to account for dipole fluctuations of QM region and the subsequent response in the MM region that give rise to attractive dispersion interactions. The charge+dipole ESPF-DRF method overestimates the dispersion interaction strength at long range, as has been shown previously,\cite{angyan_are_1990,liu_conical_2022} and it is expected the perform better as the ionisation energy of molecules in the MM region becomes larger.\cite{angyan_are_1990} The simple explanation for this overestimation is that the DRF method assumes an instantaneous response, which is equivalent to assuming that the energy scale of the MM region electronic excitations is much larger than that of the QM system.\cite{angyan_are_1990} \hlt{The overall accuracy in this case, and the \ce{Na+ + He} case above may therefore be partially due to a cancellation of error between the over-repulsive Pauli repulsion model and the over-attractive dispersion in the DRF model.}
	Overall, considering the simplicity of the DRF and Pauli repulsion models used here and the relatively small interaction energies (all less than \qty{1}{kcal.mol^{-1}}), we find the semi-quantitative accuracy of the ESPF-DRF method very encouraging. 
	
%	\hltt{In the Supporting Information we also show a comparison between the IEDRF results presented in Ref.~\onlinecite{liu_conical_2022} and our ESPF-DRF method for \ce{Na^{+}/Na^{0} + Xe} interaction energies, where ESPF-DRF performs reasonably well at reproducing the IEDRF results in these tricky examples with a highly polarisable Xe atom and a very diffuse valence orbital for \ce{Na^0}, although it is not quantitatively accurate. These examples show that whilst the ESPF-DRF approximation does not always quantitatively reproduce IEDRF, the error in these examples is comparable to the error expected in general in the DRF approach, and given the significantly reduced scaling of ESPF-DRF calculations with the MM system size, it is likely an acceptable approximation in many applications.  }
	
	\subsubsection{Molecule-atom interaction energies}
	
	We now consider an example of a simple molecule \ce{CH4} interacting with a single \ce{Ar} atom, where the \ce{CH4} molecule is treated as the QM system and \ce{Ar} is treated as the polarizable MM system. The highly symmetric \ce{CH4} molecule possesses no permanent dipole or quadrupole moment, so the interaction with the Ar atom is dominated by $1/R^6$ dispersion interactions at long range.
	
	\begin{figure}[h!]
		\centering
		\includegraphics[width=1\linewidth]{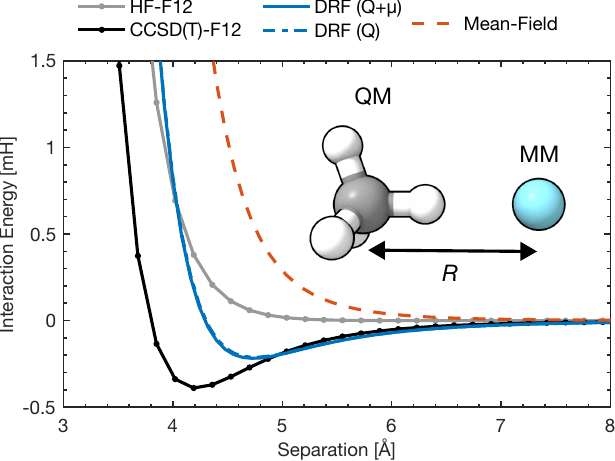}
		\caption{A) \ce{CH4 + Ar} potential energy curves calculated with HF-F12/cc-pVTZ-F12\cite{adler_simple_2007,pavosevic_geminal-spanning_2014} (grey solid line), CCSD(T)-F12/cc-pVTZ-F12 (black solid like), ESPF-DRF with charge and dipole operators (blue solid line) and just charge operators (blue dot-dash line) and ESPF mean-field theory (red dashed line) all using using PBE0/cc-pVTZ-F12. The \ce{CH4} molecule is oriented relative to the \ce{Ar} atom as shown in the insert, with one of the \ce{C-H} bonds and the \ce{C-Ar} vector beign co-linear. The $x$ axis corresponds to the \ce{C-Ar} separation. The \ce{C-H} bond lengths are set to \qty{2}{a_0}. }
		\label{fig-CH4-Ar}
	\end{figure}
	
	Because the methane-argon interaction is controlled by dispersion interactions (as is confirmed by the reference HF-F12 interaction curve being repulsive in Fig. \ref{fig-CH4-Ar}), the ESPF-MF method fails to capture the attractive interaction between the QM and MM regions, giving rise to a purely repulsive interaction. In contrast the ESPF-DRF methods, using either just charge operators or charge and dipole operators, both capture the dispersion interaction fairly accurately. Because of the 5 point charges in the QM charge expansion, the fluctuations in the dipole moment of the molecule can be captured with the charge operator model [DRF(Q)], so both DRF(Q) and DRF(Q+$\muup$) methods predict very similar interaction energies at large separations. The overestimation of the strength of dispersion interactions with the ESPF-DRF method is again expected, particularly since the QM region, \ce{CH4}, and MM region, \ce{Ar}, have similar ionisation energies, at \qty{14}{eV} and \qty{16}{eV} respectively.\cite{angyan_are_1990} %\textcolor{blue}{NF: this last sentence is not clear, \ce{Ar} being MM.}
	
	\subsubsection{Excited state interaction energies}\label{sec-exc-en}
	
	\begin{figure}[h!]
		\centering
		\includegraphics[width=1\linewidth]{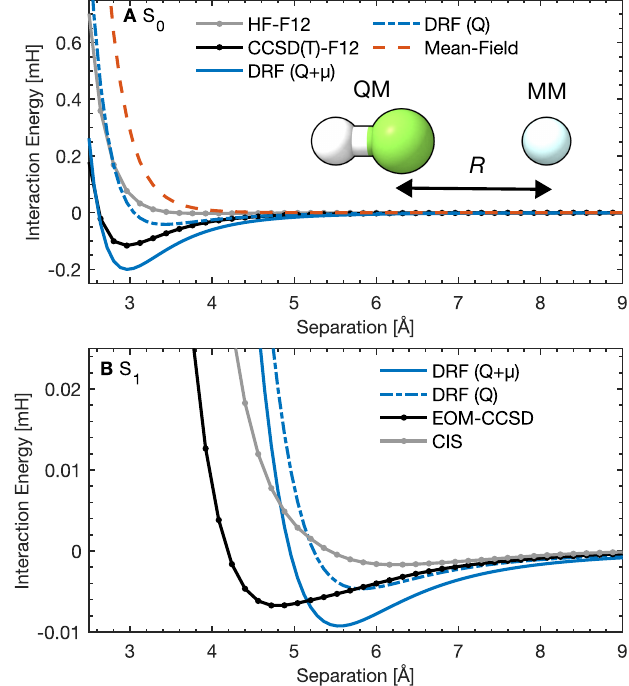}
		\caption{A) \ce{HF + He} potential energy curves in the \ce{S_1} state calculated with HF-F12/cc-pVQZ-F12 (grey solid line), CCSD(T)-F12/cc-pVQZ-F12 (black solid like), ESPF-DRF with charge and dipole operators (blue solid line) and just charge operators (blue dot-dash line) and ESPF mean-field theory (red dashed line) all using using PBE0/cc-pVQZ-F12. B) The same as A but for the excited \ce{S_1} state, where the black solid line now corresponds to the EOM-CCSD/cc-pVQZ-F12 curve and the grey line to the CIS/cc-pVQZ-F12 curve. The ESPF-DRF results now correspond to TDA-TDDFT at the PBE0/cc-pVQZ-F12 level of theory.}
		\label{fig-HF-He}
	\end{figure}
	
	Having demonstrated that the ESPF-DRF method qualitatively captures the correct physics of intermolecular interactions between QM and MM subsystems for ground-states, and predicts their interaction energies semi-quantitatively, we move on to consider excited state interaction energies. We first consider the interaction of a \ce{HF} molecule (the QM subsystem) with a \ce{He} atom (the MM subsystem), in both the ${}^1\Sigma^+$ ground state, denoted \ce{S_0}, and the ${}^1\Pi$ first excited singlet state, denoted \ce{S_1}. The ${}^1\Pi$ \ce{S_1} state correspond to a $\ce{F}\!\  2\mathrm{p}\!\to\!\sigma^*$ excitation from the closed-shell \ce{S_0} ground state , and as such there is a large change in \ce{HF} dipole moment between the two electronic states, \qty{-0.70}{a_0.e} (\qty{-1.8}{Debye}) in \ce{S_0} and +\qty{1.5}{a_0.e} (+\qty{3.8}{Debye}) in \ce{S_1} (unrelaxed EOM-CCSD dipole moments). Based on this dipole change, one might expect the \ce{S_1} state to bind more strongly with the \ce{He} atom due to the larger dipole-induced dipole interaction in the \ce{S_1} state, but in fact the binding energy is over a factor of 25 smaller for \ce{S_1} state as is seen in Fig.~\ref{fig-HF-He} (note the different energy axis scales between the \ce{S_0} interaction energies in Fig.~\ref{fig-HF-He}A and the \ce{S_1} interaction energy in Fig.~\ref{fig-HF-He}B). This is due to a significant difference in dispersion interactions in \ce{S_0} and \ce{S_1}, as is evident from the difference between the Hartree-Fock and CCSD(T)-F12\cite{adler_simple_2007,pavosevic_geminal-spanning_2014} interaction curves in \ce{S_0} and the CIS\cite{foresman_toward_1992} and EOM-CCSD\cite{koch_coupled_1990} interaction curves in \ce{S_1}, as shown in Fig.~\ref{fig-HF-He}A and B. \hltt{This is likely due to the contraction of the F atom orbitals when charge is transferred from F to H in the \ce{S_1} state, reducing the magnitude of dipole fluctuations on this atom, and therefore reducing the strength of dispersion interactions.} %Note that the good agreement between the MP2/aug-cc-pVQZ and CCSD(T)-F12/cc-pVQZ-F12 curves in Fig.~\ref{fig-HF-He}A suggests that the CIS(D) method, which is the excited state analogue of MP2, provides a reliable estimate for the interaction energies in the excited state.
	
	Now let us examine the accuracy of the QM/MM approaches. For the \ce{S_0} interaction curve, Fig.~\ref{fig-HF-He}A, the ESPF-DRF method with dipole operators [DRF(Q+$\muup$)] performs well, but as generally expected\cite{angyan_are_1990} %\textcolor{blue}{NF: why is it generally expected? }
	it overestimates the strength of dispersion interactions. The charge operator ESPF-DRF method [DRF(Q)] underestimates the interaction because it can only capture dipole fluctuations along the HF axis, so dispersion contributions from dipole fluctuations orthogonal to the \ce{H-F} bond are not captured. The ESPF-MF method massively underestimates the interaction strength, because it ignores dispersion effects, which even for a highly polar molecule like \ce{HF} dominate the intermolecular interaction. For the \ce{S_1} interaction curve, Fig.~\ref{fig-HF-He}B, we see that the ESPF-DRF methods with TDA-TDDFT\cite{hirata_time-dependent_1999} with the PBE0 functional capture the reduction in the binding energy by factor of $\sim\! 20$ relative to the \ce{S_0} state (note the change in energy axis scale between Fig.~\ref{fig-HF-He}A and B), as well as the significant increase in equilibrium separation between HF and He. We emphasise again that the \ce{S_1} state of \ce{HF} has a larger dipole moment than the \ce{S_0} state, so naively one might expect this state would bind more strongly with \ce{He}. However the difference in dispersion interactions dominates. We also note that whilst one could add an empirical dispersion correction to the \ce{S_0} interaction in order to capture the \ce{HF + He} interaction, this would clearly not be transferable to the excited \ce{S_1} state, and therefore a method like the ESPF-DRF approach which automatically includes state-specific dispersion interactions is the only way to capture, even qualitatively, the different interactions in different electronic states, and the large changes in interaction strength and equilibrium separation that result from this. 
	
	\begin{figure*}[t]
		\centering
		\includegraphics[width=1\linewidth]{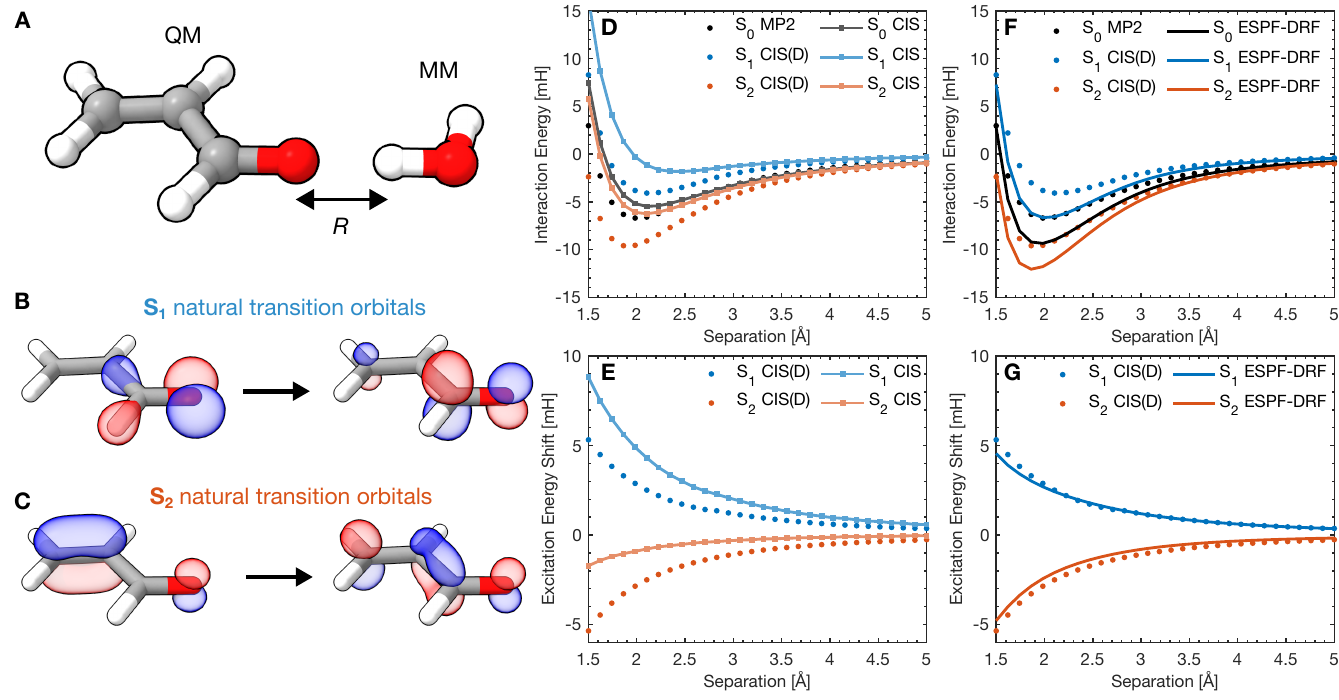}
		\caption{A) The geometry of the acrolein + water system with the \ce{C=O-HOH} separation which is scanned shown. B) Highest amplitude natural transition orbitals for the \ce{S_0}$\to$\ce{S_1} excitation of acrolein at the CIS(D)/aug-cc-pVQZ level of theory. C) same as F for the \ce{S_0}$\to$\ce{S_2} transition. 
			D) Ground and excited state potential energy curves for acrolein+\ce{H2O} calculated using MP2/CIS(D) (dots) and HF/CIS (lines and squares) with the aug-cc-pVQZ basis as a function of \ce{O-H} separation as shown in panel E. E) Excitation energy shifts (as described in the main text) as a function of separation for MP2/CIS(D) and HF/CIS. F) Same as D but comparing MP2/CIS(D) with ESPF-DRF $\omega$B97X-D3/def2-TZVP results for the excited state potential energy curves. G) Same as E but with ESPF-DRF shifts. }
		\label{fig-Acrolein-H2O}
	\end{figure*}
	
	% \textcolor{blue}{NF: there is no mention on how the equilibrium distance between \ce{HF} and \ce{Ar} changes with the method. The charge to charge+dipole improvement is much better in \ce{S_0} than in \ce{S_1}. Naively, we may think that the addition of ESPF quadrupoles could improve it (ie reduce it toward the EOM-CCSD value).}
	
	As a final example we consider the energy of the ground and first two excited states of acrolein interacting with a single water molecule. The MP2/CIS(D) method\cite{head-gordon_doubles_1994} is used together with the aug-cc-pVQZ basis set to obtain energies of acrolein in the \ce{S_0}, \ce{S_1} and \ce{S_2} states interacting with a \ce{H2O} molecule, with the \ce{C=O} and one of the water \ce{H-O} bonds co-linear, as shown in Fig.~\ref{fig-Acrolein-H2O}A. The \ce{S_1} state arises from a $\mathrm{n}\!\to\!\pi^*$ excitation from a non-bonding \ce{O} lone-pair orbital to the \ce{C=O} $\pi^*$ orbital, as shown by the natural transition orbitals in Fig.~\ref{fig-Acrolein-H2O}B between the \ce{S_0} and \ce{S_1} states, and the \ce{S_2} state corresponds to a $\pi\!\to\!\pi^*$ excited state, as shown by the natural transition orbitals in Fig.~\ref{fig-Acrolein-H2O}C. %\textcolor{blue}{NF: it would be more logical to have this 3 pictures as A, B and C in Figure 4. }
	In the \ce{S_1} state, electron density is transferred away from the O atom, and as a result the stabilising electrostatic interaction with the polar \ce{H-O} bond is reduced and the \ce{S_1} state binds less strongly to the \ce{H2O}. In contrast in the \ce{S_2} state, electron density is transferred from the \ce{C=C} bond to the \ce{C=O} bond, so there is a stronger electrostatic interaction with the polar \ce{H-O} bond, and the \ce{S_2} state binds more strongly to \ce{H2O} molecule. Dispersion and electron correlation effects also play a significant role, as seen by comparing the HF/CIS interaction energies for each state (which neglect electron correlation and dispersion) to the MP2/CIS(D) interaction curves, Fig.~\ref{fig-Acrolein-H2O}A. Dispersion interactions stabilise all three states, but more significantly in the excited states. 
	
	For this system we have also calculated the interaction energies using ESPF-DRF with DFT using the $\omega$B97X-D3 functional\cite{lin_long-range_2013} and TDDFT for the excited states (with the Tamm-Dancoff approximation\cite{hirata_time-dependent_1999}) and the def2-TZVP basis set as shown in Fig.~\ref{fig-Acrolein-H2O}F [note that the D3 correction is only included within the QM sub-system, in this case the acrolein molecule]. The water model parameters correspond to the ``Dipole 1'' model in Table \ref{tab-model-params-dip}. Overall the positions of the minima for each excited state are captured well using ESPF-DRF, although the binding energies are overestimated slightly, due to DRF in general overestimating the dispersion interaction strength. The significant stabilisation of the excited states by dispersion is captured semi-quantitatively by the ESPF-DRF method. %\textcolor{blue}{NF: is the previous sentence necessary here?} 
	Later we will consider the application of ESPF-DRF to calculate solvatochromic shifts, where the measured quantity is the shift in excitation on interaction with a solvent, and so with this in mind in Fig.~\ref{fig-Acrolein-H2O}B and D we show the excitation energy shift, $\Delta\Delta E_{\ce{S}_n}(R) = \Delta E_{\ce{S}_n}(R) - \Delta E_{\ce{S}_n}(R=\infty)$, where $\Delta E_{\ce{S}_n}(R) = E_{\ce{S}_n}(R) - E_{\ce{S}_0}(R)$ as a function of the separation of acrolein and the \ce{H2O} molecule, calculated using the CIS(D) method, CIS [Fig.~\ref{fig-Acrolein-H2O}B] and ESPF-DRF [Fig.~\ref{fig-Acrolein-H2O}D]. We see that ESPF-DRF very accurately captures the shift for the \ce{S_1} state, whereas the CIS method, which neglects dispersion energy differences, considerably overestimates the shift. The ESPF-DRF method underestimates the shift for the \ce{S_2} state but it still performs better than CIS for the full acrolein+\ce{H2O} system. Even though the ESPF-DRF method may overestimate the total interaction energy between acrolein and \ce{H2O} for the three different electronic states, this error is similar for each state at a given separation, so these errors cancel for the excitation energy shift as a function of separation, which is captured very accurately.
	
	\subsection{Solvatochromic shifts}
	
	% \begin{SCfigure*}[0.7][t]
		%     \includegraphics[width=0.68\textwidth]{fig-acrolein-spec.pdf}
		%     \caption{}
		%     \label{fig-acrolein-solv}
		% \end{SCfigure*}

	\begin{figure*}[t]
		\includegraphics[width=1.0\textwidth]{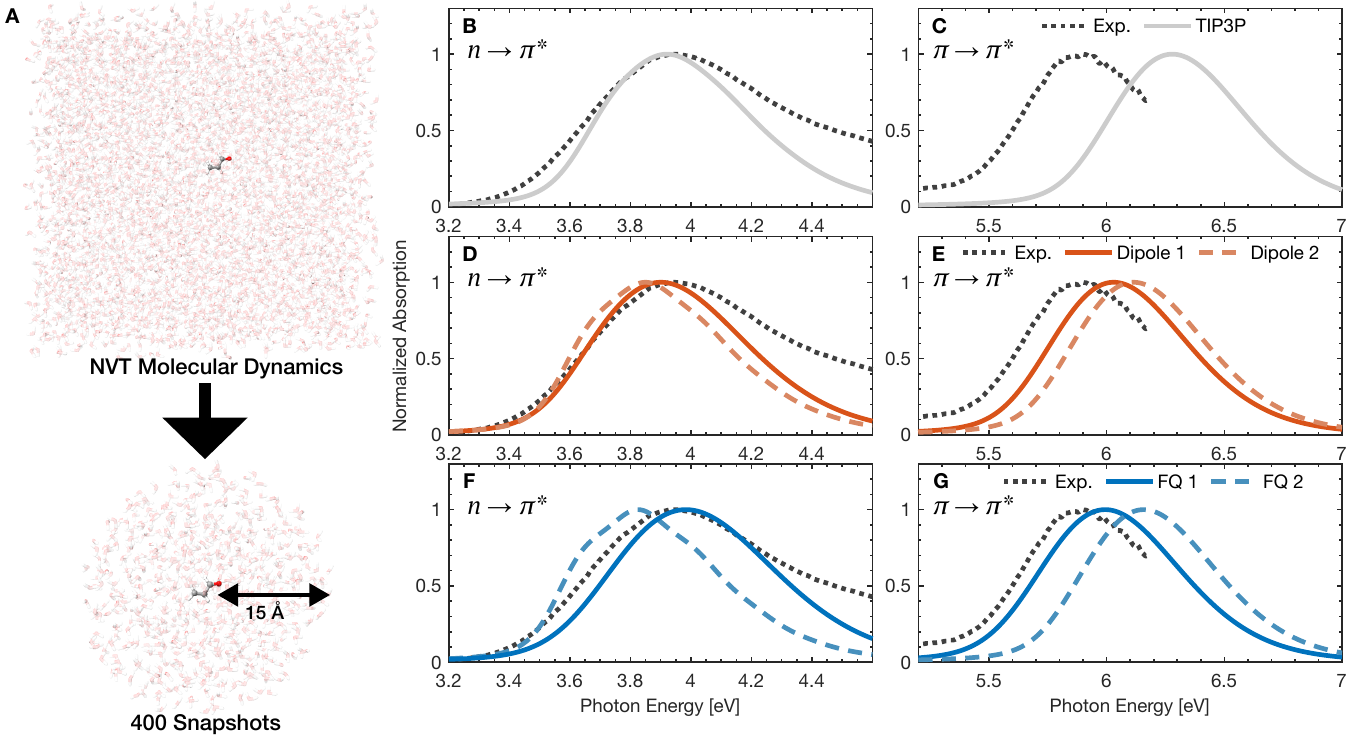}
		\caption{A) An illustration of the workflow used to sample configurations for the acrolein spectrum simulations. B) Experimental and calculated absorption spectra for the $n\to\pi^*$ transition in acrolein using the TIP3P fixed charge model for \ce{H2O}. C) Same as B for the $\pi\to\pi^*$ transition. D) Experimental and calculated absorption spectra for the $n\to\pi^*$ using ESPF-DRF with dipole polarizability models for \ce{H2O}. E) Same as D for the $\pi\to\pi^*$ absorption spectrum. F) Experimental and calculated absorption spectra for the $n\to\pi^*$ using ESPF-DRF with fluctuating charge models for \ce{H2O}. G) Same as F for the $\pi\to\pi^*$ transition. }
		\label{fig-acrolein-solv}
	\end{figure*}
	
	As a further test for the ESPF-DRF method, we have calculated gas to aqueous phase solvatochromic shifts for the first two optical absorption bands in acrolein, where we can compare directly to experimental absorption spectra.\cite{aidas_performance_2008,lee_substituent_2007} Full details of the absorption spectrum calculations are given in Appendix \ref{app-spec-calc}, but we briefly summarise the procedure here before showing results. Firstly, configurations for the acrolein in the gas phase and in aqueous solution were sampled using molecular dynamics, with the TIP3P water model and a bespoke force field parameterized for acrolein, using the procedure described in Ref.~\onlinecite{fay_unraveling_2024}. Overall 400 configurations were sampled every \qty{0.25}{ns} from a \qty{100}{ns} molecular dynamics simulation with a Langevin thermostat to maintain the temperature at \qty{298}{K}, all performed with OpenMM 8.\cite{eastman_openmm_2017,eastman_openmm_2024} From each of these snapshots a droplet of water was extracted by selecting all water molecules within \qty{15}{\AA} of any acrolein atom using MDAnalysis,\cite{michaudagrawal_mdanalysis_2011,gowers_mdanalysis_2016} and ESPF-DRF calculations were run with TDDFT with the $\omega$B97X-D3 functional and def2-TZVP basis set (this workflow is illustrated in Fig.~\ref{fig-acrolein-solv}A). These results are combined to produce a classical static disorder approximation for the spectrum, which neglects vibronic effects in the spectrum.\cite{zuehlsdorff_combining_2018,zuehlsdorff_optical_2019} In order to account for this the final spectra are calculated using a displaced harmonic oscillator model (i.e. a spin-boson mapping), parameterized to reproduce the band maximum and line-width from the static disorder spectrum, with Huang-Rhys factors for the intramolecular reorganisation parameterized from gas phase electronic structure calculations.\cite{zuehlsdorff_combining_2018,zuehlsdorff_optical_2019,segarra-marti_modeling_2020} The details of this and the approximations made are given in Appendix \ref{app-spec-calc}, and validation of the method by comparison to gas phase spectra is given in Appendix \ref{app-gas-spec}.
	
	In most of the previous examples we considered simple monatomic MM systems, where the polarizability can be fitted using a high-level \textit{ab initio} calculation. For a molecular system like \ce{H2O} charges and polarizabilities (or in the case of fluctuating charge models, electronegativities $\chi_i$ and chemical hardness parameters $\eta_i$) model parameters can be parameterized in multiple ways.\cite{nicoli_assessing_2022} Several dipole polarizability (denoted Dipole 1 and Dipole 2) and fluctuating charge models (denoted FQ 1 and FQ 2) have been proposed for \ce{H2O}, and here we test four of them combined with the ESPF-DRF framework, as well as a simple fixed charge TIP3P model, which does not include polarizability in any way.\cite{nicoli_assessing_2022} These model parameters are summarised in Tables \ref{tab-model-params-dip} and \ref{tab-model-params-fq}.
	
	In Fig.~\ref{fig-acrolein-solv} we show the simulated aqueous solution spectra calculated using a range of models for the water molecules in the MM region, and Table~\ref{tab-acr-spec} summarises the peak positions and solvatochromic shifts predicted by the different methods. Experimental spectra are taken from Refs.~\onlinecite{aidas_performance_2008} and \onlinecite{lee_substituent_2007}. The calculated $n\!\to\!\pi^*$ (Fig.~\ref{fig-acrolein-solv}B,D,F) have been shifted by \qty{-0.06}{eV}, based on the error in the gas phase peak positions relative to experiment. The systematic underestimation of the predicted absorption line width for this transition can be attributed to the chosen electronic structure method, TDDFT $\omega$B97X-D3/def2-TZVP, underestimating the intramolecular reorganisation energy (see Appendix \ref{app-gas-spec} for more details). Starting with the non-polarizable TIP3P model for \ce{H2O}, we see that this fixed charge model gives a reasonably good estimate for the blue-shift in the $n\!\to\!\pi^*$ transition (Fig.~\ref{fig-acrolein-solv}B). The polarizable models all predict qualitatively correct shifts for the $n\!\to\!\pi^*$ transition (Fig.~\ref{fig-acrolein-solv}D and F), although the quality varies with the model. The best performers are the fluctuating charge ``FQ 1'' model and the``Dipole 1'' dipole-polarizability model, both of which have errors of $<15\%$ in the shift. The poor performance of the ``Dipole 2'' model can likely be attributed to the fact that it treats the water molecule with a single polarizable site on the oxygen atom, so effects arising from polarization on the H atoms in water, as is likely important in hydrogen bonding between water and the \ce{C=O} bond in acrolein, cannot be accounted for properly with the ``Dipole 2'' model. The failure of the ``FQ 2'' fluctuating charge model likely originates in the higher hardness parameter, $\eta_\mathrm{H}$, of H which reduces the size of the charge fluctuations on the H atoms, which again is likely to be important due to hydrogen bonding with the \ce{C=O} group on acrolein.
	
	These same models that perform best for the $n\to\pi^*$ transition also perform the best in predicting the red shift in the $\pi\!\to\!\pi^*$ transition, although for this transition the Dipole 1 model underestimates the shift by about 28\% and the FQ 1 model underestimates it by 20\%. It should also be noted that the TIP3P non-polarizable model significantly underestimates the red shift of the $\pi\!\to\!\pi^*$ transition (Fig.~\ref{fig-acrolein-solv}C). Environment polarization and differences in dispersion interactions are clearly play a significant role in determining the magnitude of solvatochromic shifts for these transitions, especially for the $\pi\!\to\!\pi^*$ which involves a more significant degree of charge transfer. Capturing these effects accurately however requires an accurately parameterized model for the solvent polarizability, as is shown by the range of shifts obtained with different models for \ce{H2O} electrostatics and polarizability. The results shown would certainly be sensitive to the solvent and solute force fields (particular non-bonded parameters) used in sampling configurations. Our aim here is not to provide a rigorous assessment of different polarizable models, although the performance of the models tested here agrees with more thorough benchmarking which others have performed\cite{nicoli_assessing_2022} using the mean-field linear-response framework for polarizable QM/MM, but rather to demonstrate that models for the MM environment which have already been parameterized using other methods can be transferred to predict solvation effects on excitation energy within the ESPF-DRF framework.
	\begin{SCtable*}[0.7][t]
		\centering
		\begin{tabular}{ccccc}
			\hline
			Method & $E_\mathrm{max}(\mathrm{n}\!\to\!\pi^*)$ & $\Delta E_\mathrm{max}(\mathrm{n}\!\to\!\pi^*)$ & $E_\mathrm{max}(\pi\!\to\!\pi^*)$ & $\Delta E_\mathrm{max}(\pi\!\to\!\pi^*)$ \\
			\hline  
			Experiment (Solution)${}^a$  & 3.94 & +0.25 & 5.89 & $-$0.52 \\
			Experiment (Gas)$^{b}$ & 3.69 & -- & 6.41 & -- \\\
			Gas & 3.752$\pm0.016$ & -- & 6.405$\pm0.028$ & -- \\
			Gas (Re-weighted)$^{c}$ & 3.772$\pm0.020$ & -- & 6.437$\pm0.036$ & -- \\
			Fixed charge (TIP3P) & 3.980$\pm0.026$ & +0.228$\pm0.032$ & 6.279$\pm0.032$ & $-0.127\pm0.043$ \\
			ESPF-DRF (Dipole 1) & 3.960$\pm0.030$ & +0.208$\pm0.034$ & 6.031$\pm0.034$ & $-0.374 \pm0.044$ \\
			ESPF-DRF (Dipole 2) &   3.910$\pm0.024$ & +0.160$\pm0.029$ & 6.115$\pm0.035$ & $-0.290 \pm0.045$\\
			ESPF-DRF (FQ 1) & 4.044$\pm0.042$ & +0.293$\pm0.045$ & 5.993$\pm0.036$ & $-0.412 \pm0.046$\\
			ESPF-DRF (FQ 2) & 3.886$\pm0.021$ & +0.134$\pm0.026$ & 6.157$\pm0.032$ & $-0.248 \pm0.043$ \\
			\hline
		\end{tabular}
		\caption{Summary of experimental and calculated maximum absorption data for the $n\to\pi^*$ and $\pi\to\pi^*$ transitions in acrolein including positions of the maximum absorption $E_\mathrm{max}$ and solvatochromic shifts $\Delta E_{\mathrm{max}} = E_\mathrm{max,aq} - E_\mathrm{max,gas}$. ${}^a$Data from Ref.~\onlinecite{aidas_performance_2008}. ${}^b$Data from Ref.~\onlinecite{aidas_performance_2008} and \onlinecite{lee_substituent_2007} ${}^c$From thermal re-weighting of configurations sampled from MM force field based MD using calculated QM energies.  }
		\label{tab-acr-spec}
	\end{SCtable*}

	\section{Conclusions}
	
	In this work we have outlined the ESPF-DRF method for incorporating polarization and dispersion effects into QM/MM energy calculations. The method makes use of an atom-centred multipole expansion of the QM region charge density to yield an efficient method for calculating direct reaction field polarization energies, avoiding the expensive computation of large matrix inverses. The ESPF-DRF method side-steps these large matrix inverses, avoiding the $\mathcal{O}(N_\MM^3)$ cost and $\mathcal{O}(N_\MM^2)$ memory requirement of the DRF method, improving the \hlt{efficienct of the method, particularly for large MM systems}. Because the ESPF-DRF method does not involve any additional electron integrals beyond those commonly available in standard electron integral packages, it can be straightforwardly implemented in many existing codes. Our PySCF add-on demonstrates how this can be achieved relatively simply.\cite{pyespf} Although the use of ESPF multipoles is approximate compared to the integral exact formulation,\cite{liu_conical_2022,humeniuk_multistate_2024} comparison with high-level electronic structure calculations shows that it is accurate even for very weak interaction energies. Overall the ESPF-DRF method can capture state-specific polarization effects, which are difficult and very costly to capture with mean-field polarizable QM/MM methods or simple fixed-charge models, as well as state-specific dispersion effects, which simple empirical pair-wise dispersion corrections\cite{giovannini_general_2017} cannot capture. The accuracy of the method has been tested in a range of systems, including for electronic excited states, and its utility has been demonstrated in calculating accurate gas to aqueous solution solvatochromic shifts for acrolein.
	
	Looking to the future, we anticipate the ESPF-DRF will be a useful method for exploring spectroscopy and excited state processes in condensed phase environments. Analytic gradient and hessian calculations, whilst not yet implemented, should be straightforward to develop using existing theoretical frameworks.\cite{bonfrate_analytic_2024,schwinn_analytic_2019,huix-rotllant_analytic_2021} There are cases where the ESPF approximation may breakdown, namely when very short range interactions are important, and for such cases it may be possible to combine ESPF-DRF for long-range interactions with the exact IEDRF method for short-range interactions, yielding a mixed method with the advantages of both approaches. The ESPF-DRF method is also directly compatible with modern machine learning force-fields which use fluctuating charge (also known as charge equilibration) schemes to treat long-range electrostatics,\cite{gubler_accelerating_2024,ko_accurate_2023} which will enable the development of QM/ML approaches with ESPF-DRF. In the immediate term we anticipate that ESPF-DRF will provide a useful tool for investigating optical properties of molecules in solution, such as absorption/fluorescence properties, solvatochromism and circular dichroism in complex condensed phase environments, such as in solvents, proteins and at interfaces. In order to tackle these problems we hope that our publicly available add-on to PySCF will help by enabling access to the DRF methodology through open source software.
	
	% \begin{enumerate}
		%     \item Good overall performance without system-specific parameterization.
		%     \item Scalable for large systems.
		%     \item Compatibility with modern ML models for long-range electrostatics.
		% \end{enumerate}
	
	% \subsection*{Author contributions}
	
	\subsection*{Acknowledgements}
	\vspace{-10pt}
	\noindent We would like to thank William Glover for providing data and simulation parameters for calculations from Ref.~\onlinecite{liu_conical_2022}. This work was supported by ``Agence Nationale de la Recherche'' through the project MAPPLE (ANR-22-CE29-0014-01). 
	
	\vspace{-10pt}
	\subsection*{Supporting information}
	\vspace{-10pt}
	
	\noindent Supporting information includes an analysis of the computational scaling of the methods described in this paper and IEDRF (Sec.~S.1), an analysis of the role of the corrections to the ESPF operators (Sec.~S.2), and the effect of the damping function on ESPF-DRF calcualtions (Sec.~S.3).
	
	\vspace{-10pt}
	\subsection*{Data availability}
	\vspace{-10pt}
	
	\noindent The PyESPF code, provided for free at \url{www.github.com/tomfay/PyESPF}, provides an interface for performing ESPF-DRF calculations with the open source electronic structure code PySCF. All data presented in this paper is publicly available. Data for test bimolecular systems is available at \url{www.zenodo.org/doi/10.5281/zenodo.13736077} and scripts for running the ESPF-DRF calculations are available at \url{www.github.com/tomfay/PyESPF}. Data for acrolein spectrum calculations, including force fields, OpenMM scripts, configurations, and ESPF-DRF energy data, are available at \url{www.zenodo.org/doi/10.5281/zenodo.13735909}. 
	
	\vspace{-10pt}
	\subsection*{Code availability}
	\vspace{-10pt}
	
	\noindent The PyESPF code, provided for free at \url{www.github.com/tomfay/PyESPF}, provides an interface for performing ESPF-DRF calculations with the open source electronic structure code PySCF, together with example scripts for the QM/MM calculations in this paper.
	
	\vspace{-10pt}
	\subsection*{Conflicts of interest}
	\vspace{-10pt}
	\noindent The authors declare no conflicts of interest.
	
	\appendix
	% set new figure/table numbering for appendix
	\renewcommand\thefigure{\thesection.\arabic{figure}}    
	\setcounter{figure}{0}   
	\renewcommand\thetable{\thesection.\arabic{table}}   
	\setcounter{table}{0}   
	\vspace{-10pt}
	\section{Fluctuating charge DRF}\label{app-fq-drf}
	\vspace{-10pt}
	In this appendix we expand on the explicit equations for fluctuating charge (FQ)\cite{lipparini_linear_2012,rick_dynamical_1994} DRF method. It is functionally equivalent to dipole DRF, and we just need to provide equations for the terms $E_\mathrm{FQ,0}$m $\vb*{q}_0$ and $\vb{K}_\mathrm{FQ}$ in Eq.~\eqref{eq-fq-energy}. The subtlety in the FQ method compared to dipole approach is in enforcing constraints on the total charge of the system, or individual molecules within the MM system. Firstly we define a set of $g=1,...,N_\mathrm{con}$ fixed charge groups
	\begin{align}
		Q_g = \sum_{k=1}^{N_g}q_{i_k^g}
	\end{align}
	and we assume that the fixed charge groups are disjoint. These groups are defined in the chosen model, for example in the FQ water models we have applied we enforce that each water molecule has a net 0 charge. 
	Typically charge constraints are handled by introducing Lagrange multipliers\cite{rick_dynamical_1994,lipparini_linear_2012} but because the constraint is linear in $\vb*{q}$ it can actually be handled straightforwardly by direct projection onto a subset of free charges $\vb*{q}'$. 
	The set of $N_\MM - N_\mathrm{con}$ free charges $\vb*{q}'$ is related to the full vector $\vb*{q}$ of $N_\MM$ fluctuating charges by
	\begin{align}
		\vb*{q} = \vb{P}\vb*{q}' +\vb{S} \vb{P}\vb*{q}' + \vb*{Q}
	\end{align}
	where $\vb{P}$ projects from the set of free charges to the full set, the matrix $\vb{S}$ is given by
	\begin{align}
		S_{ij} = -\sum_{g} \delta_{i,i^g_{N_g}}\sum_{k=1}^{N_g-1}\delta_{j,i^g_{k}}
	\end{align}
	and $\vb*{Q}$ is given by
	\begin{align}
		Q_i = \sum_g\delta_{i,i^g_{N_g}} Q_g.
	\end{align}
	With these definitions we can further define the following matrices
	\begin{align}
		\vb{U} &= \vb{P} + \vb{S P} \\
		\vb*{\eta}' &= \vb{U}^\intercal \vb*{\eta}\vb{U} \\
		\vb{T}' &= \vb{U}^\intercal \vb{T}^\mathrm{FQ}\vb{U} 
	\end{align}
	and $\vb*{\eta}$ is a diagonal matrix of the the chemical hardness parameters for each atom.
	
	With the above matrices in hand the terms in Eq.~\eqref{eq-fq-energy} are given by
	\begin{align}
		\begin{split}
			E_{\mathrm{FQ,0}} &= \vb*{Q}^\intercal (\vb*{\phi}_{\MM} + \vb*{\chi}) \\
			&+ \frac{1}{2}\vb*{Q}^\intercal(2\vb*{\eta}+\vb{T}^\mathrm{FQ})\vb*{Q}- \frac{1}{2}\vb*{b}_0^\intercal(2\vb*{\eta}'+\vb{T}')^{-1}\vb*{b}_0 
		\end{split} \\
		\vb*{b}_0 &= \vb{U}^\intercal (\boldsymbol{\phi}_{\MM}+\boldsymbol{\chi}) +\vb{U}^\intercal (2 \vb*{\eta} + \vb{T}^\mathrm{FQ}) \vb*{Q}
	\end{align}
	where $\vb*{\phi}_\MM$ is a vector of potentials generated by MM region fixed charges at each MM site, and $\vb*{\chi}$ is a vector of electronegativities. The term $\vb*{q}_0$ is the set if charges generated in the absence of the QM region charges,
	\begin{align}
		\vb*{q}_0 = \vb*{Q} - \vb{U}(2\vb*{\eta}' + \vb{T}')^{-1}\vb*{b}_0
	\end{align}
	The matrix $\vb{K}_\mathrm{FQ}$ is given by 
	\begin{align}
		\vb{A} = \vb{U}(2\vb*{\eta}'+\vb{T}')^{-1}\vb{U}^\intercal 
	\end{align}
	which we see requires the matrix in verse $(2\vb*{\eta}'+\vb{T}')^{-1}$. When the QM region potential $\vb*{\phi}[\hat{\rho}_\QM]$ is given by the ESPF expansion, the full matrix is not needed and we can just solve systems of linear equations instead. 
	
	Using the above, we re-define the $U_{\QM/\MM}^a$ and $U_{\QM}^{ab}$ for the FQ method as
	\begin{align}
		U_{\QM/\MM}^a &= -\vb*{q}_0^\transp \vb*{\phi}_a \\
		U_{\QM}^{ab} &= {\phi}_{a}^\transp \vb*{K}_{\mathrm{FQ}}\vb*{\phi}_b .
	\end{align}
	These, together with the replacement $\vb*{\mathcal{E}}_\MM^\transp \vb{K}_\mathrm{D}\vb*{\mathcal{E}}_\MM/2 \to E_\mathrm{FQ,0}$ in Eq.~\eqref{eq-Vpol0} fully define the ESPF-DRF method with a fluctuating charge model for the MM region polarization. The set of $N_\mathcal{Q}+1$ systems of linear equations that need to be solved to find $U_{\QM/\MM}^a$ and $U_{\QM}^{ab}$ can be solved iteratively, involving at most $\mathcal{O}(N_\mathrm{MM}^2)$ operations per iteration. Thus this approach scales more favourably that direct evaluation of the matrix inverse.
	
	As an example of the ESPF-DRF method using a fluctuating charge model (FQ 1 in Table \ref{tab-model-params-fq}) for the molecular environment we have considered the acrolein+\ce{H2O} system that we have also performed calculations on using the Dipole 1 polarizable \ce{H2O} model in Fig.~\ref{fig-Acrolein-H2O} in Sec.~\ref{sec-exc-en}. The results using the fluctuating charge model are shown in Fig.~\ref{fig-acrolein-h2o-fq1}. We see that this model underestimates the binding energy for each excited state, likely because the fluctuating charge model cannot capture polarization effects and dispersion interaction arising from dipole fluctuations perpendicular to the plane of water molecule. When considering the excitation energy shifts however the model performs well compared to the reference CIS(D) results, except for at smaller separations.
	
	\begin{figure}[ht]
		\centering
		\includegraphics[width=0.95\linewidth]{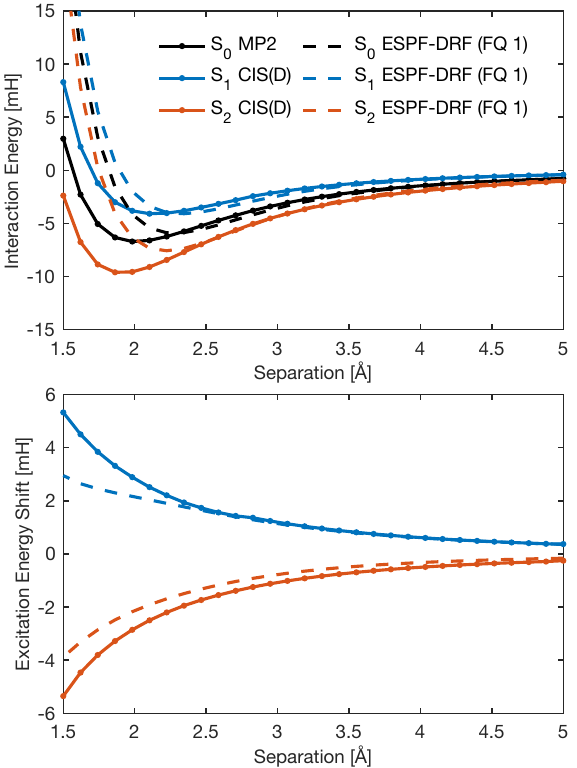}
		\caption{A) Reference and ESPF-DRF potential energy curves for the ground and first two excited states of acrolein interacting with \ce{H2O} (as in Fig.~\ref{fig-Acrolein-H2O}) using the FQ 1 model for \ce{H2O} (See Table \ref{tab-model-params-fq} for model parameters). B) Same as A but for the excitation energy shifts as a function of acrolein \ce{H2O} separation.}
		\label{fig-acrolein-h2o-fq1}
	\end{figure}
	\vspace{-10pt}
	\section{Pauli repulsion model}\label{app-pauli-rep}
	\vspace{-10pt}
	The Pauli repulsion model used here is based on the empirical observation that penetration interaction energy and exchange interaction between fragments A and B is approximately $E_{\mathrm{pen,AB}} \approx - 2 E_\mathrm{ex,AB}$, so the overall repulsion energy is approximately\cite{giovannini_general_2017,amovilli_matrix_1990,amovilli_self-consistent-field_1997}
	\begin{align}
		E_\mathrm{rep,AB} = E_\mathrm{pen,AB} + E_\mathrm{ex,AB} \approx -E_\mathrm{ex,AB}
	\end{align}
	The exchange interaction is given in terms of the one-body density matrices for A and B as
	\begin{align}
		E_\mathrm{ex,AB} \!=\! - \!\sum_{\sigma_1\sigma_2}\!\! \int\!\!{\dd{\vb*{r}_1}}\!\!\!\int\!\!{\dd{\vb*{r}_2}}\! \frac{1}{r_{12}} \Gamma_{\mathrm{A},\sigma_1\sigma_2}(\vb*{r}_1,\vb*{r}_2) \Gamma_{\mathrm{B},\sigma_2\sigma_1}(\vb*{r}_2,\vb*{r}_1)
	\end{align}
	If the B fragment is closed shell then $\Gamma_{\mathrm{B},\sigma_2\sigma_1}(\vb*{r}_2,\vb*{r}_1) = (1/2)\delta_{\sigma_1\sigma_2}\Gamma_\mathrm{B}(\vb*{r}_2,\vb*{r}_1)$, so
	\begin{align}
		E_\mathrm{ex,AB} \!= \!-\frac{1}{2}  \int\!\!{\dd{\vb*{r}_1}}\!\!\int\!\!{\dd{\vb*{r}_2}}\frac{1}{r_{12}} \Gamma_{\mathrm{A}}(\vb*{r}_1,\vb*{r}_2) \Gamma_{\mathrm{B}}(\vb*{r}_2,\vb*{r}_1).
	\end{align}
	In order to use this model for Pauli repulsion in QM/MM calculations we just need a model for the MM region one-body reduced density. This can be done in many ways, for example by directly parameterizing a model to high-level calculations, but instead we use a simple approach, where we assume the exchange-repulsion is dominated by electrons in the highest energy sub-shell of each MM atom, so we approximate the reduced density matrix with a sum of atom-centred pseudo orbitals
	\begin{align}
		\Gamma_{\MM}(\vb*{r}_1,\vb*{r}_2) \approx \sum_A N_A \varphi_A(\vb*{r}_1)\varphi_A(\vb*{r}_2).
	\end{align}
	$N_A$ is set to the number of electrons in the highest energy sub-shell of the neutral atom plus any excess from the MM model charges i.e. $N_A = N_A^\mathrm{atom} - q_A$. The pseudo-orbitals $\varphi_A(\vb*{r})$ are parameterized to reproduce the highest energy sub-shell electron density based on van der Waals radii of the atoms.\cite{rahm_atomic_2016} In our case we use a re-scaled STO-3G type orbital\cite{dunning_gaussian_1989,pritchard_new_2019} for each of these pseudo-orbitals. The orbital coefficient is fitted such that
	\begin{align}
		|\varphi(\vb*{R}_A + R^* \vb*{e})|^2 = 8\pi \alpha^3 e^{-\alpha R^*}
	\end{align}
	where $\alpha$ is fitted, such that $8\pi N_A\alpha^3 e^{-\alpha R_{\mathrm{vdW},A}} = 0.001$ and $R^* = 0.75 R_{\mathrm{vdW},A}$. The additional fitting step is required because the STO-3G type orbital decays too rapidly in the low density tail. This simple procedure is likely not optimal, but it is physically motivated and uses simple atomic density-derived parameters which are already readily available, so no additional parameterization is needed. 
	\vspace{-10pt}
	\section{Spectrum calculations}\label{app-spec-calc}
	\vspace{-10pt}
	\subsection{Details of spectrum calculations}
	\vspace{-10pt}
	Our starting point for calculating the optical absorption spectrum is the classical static disorder approximation,\cite{zuehlsdorff_optical_2019} which is obtained as
	\begin{align}\label{eq-spec}
		A_{\mathrm{SA}}(\omega) \!\propto\! \!\int \!\dd{\vb*{R}}\mu_{i\to f}^2(\vb*{R})\delta(\omega \!-\! \Omega_{i\to f}(\vb*{R})) e^{-\beta  E_{i}(\vb*{R})}
	\end{align}
	where $i$ denotes the initial electronic state and $f$ labels the set of excited states, $\mu_{i\to f}^2(\vb*{R})$ is the square of the transition dipole moment for the $i \to f$ excitation, which is a function of the nuclear configuration $\vb*{R}$, $\Omega_{i\to f}(\vb*{R})$ is the corresponding excitation energy as a function of the nuclear configuration $\hbar \Omega_{i\to f}(\vb*{R}) = E_f(\vb*{R}) - E_i(\vb*{R})$, and  $E_n(\vb*{R})$ is the electronic potential energy surface for state $n$. $\beta = 1/k_\mathrm{B}T$ is the reciprocal of thermal energy. The energies are approximated using the ESPF-DRF method described above. The $\delta$ function in Eq.~\eqref{eq-spec} is replaced with a Gaussian $\delta(x)\to\tilde{\delta}(x) = \sqrt{1/2\pi\Delta^2}\exp(-x^2/2\Delta^2)$ to account for missing broadening effects, with the broadening parameter $\Delta$ set to $\Delta=\qty{0.15}{eV}$.
	
	We sample configurations $\vb*{R}$ using a simple molecular mechanics force field for the initial state $U_{i}^{\MM}(\vb*{R})$ for the ground electronic state. This accelerates sampling of uncorrelated configurations significantly over using QM/MM based sampling, but at the cost of accuracy. For this purpose we have parameterized a simple force-field for acrolein in its ground electronic state, using OPLS-AA parameters\cite{jorgensen_development_1996,dodda_ligpargen_2017} to describe non-bonded interactions, following the procedure from Ref.~\onlinecite{fay_unraveling_2024}. The reference \ce{S_0} state geometry and hessian needed for the force field fitting was obtained from PBE0/def2-TZVP geometry optimisation. Charges are taken as a 50:50 mix of CHELPG charges derived from vacuum and CPCM calculations. The solution phase box was prepared using OpenMM with 5000 water molecules described with the TIP3P force field. An initial NPT equilibration was performed to determine the box volume, which was set to \qty{5.29369}{nm}$\times$\qty{5.29369}{nm}$\times$\qty{5.29369}{nm}, and the box was then equilibrated at \qty{298}{K} for \qty{1}{ns} using a Langevin thermostat with a \qty{1}{fs} time-step and a friction coefficient of \qty{2}{ps^{-1}}. In both the gas and solution phase simulations case \qty{100}{ns} of sampling is performed with configurations sampled every \qty{0.25}{ns}. In order to evaluate the accuracy of this force-field in the gas phase, we have also calculated the spectra by re-weighting gas phase configurations using the exact QM energies. In this case the re-weighting factor for sampled configuration $k$ is $w_k = \exp(-\beta (E_{i,\mathrm{QM}}(\vb*{R}_k)-E_{i,\mathrm{MM}}(\vb*{R}_k)))$.
	
	The static disorder spectrum fails to account for vibronic effects in the spectrum. In order to correct for some of the missing vibronic effects, we calculate the spectrum with an approximate Gaussian Condon  approach which is based on a displaced harmonic oscillator model (otherwise referred to as a cumulant expansion theory or spin-boson mapping)\cite{zuehlsdorff_combining_2018,fay_unraveling_2024,wiethorn_beyond_2023}
	\begin{align}
		A_{\mathrm{GC}}(\omega) \propto \frac{1}{2\pi}\int_{-\infty}^\infty \dd{t} e^{i\omega t} e^{g(t)} e^{-t/\tau}
	\end{align}
	the function $g(t)$ is decomposed as\cite{zuehlsdorff_combining_2018,fay_unraveling_2024}
	\begin{align}
		g(t) = g_\mathrm{mol}(t) + g_\mathrm{env}(t) - i \epsilon t/\hbar.
	\end{align}
	The intramolecular term $g_\mathrm{mol}(t)$ is obtained from a displaced Harmonic oscillator model parameterized from gas phase calculations of the ground and excited state equilibrium geometry and the hessian of the excited state,
	\begin{align}
		\begin{split}
			g_\mathrm{mol}(t) &= -(\lambda_\mathrm{mol}/\hbar) \times\\
			\sum_{\alpha} \!\frac{w_\alpha}{\omega_\alpha}\!&\bigg(\!\coth(\!\frac{\hbar \beta\omega_\alpha}{2}\!)(1\!-\!\cos(\omega_\alpha t))
			\!+\! i \sin(\omega_\alpha t))\!\bigg)
		\end{split}
	\end{align}
	where $\omega_\alpha$ are the normal mode frequencies of the excited state and the weight factors $w_\alpha = \Delta Q_\alpha^2 \omega_\alpha^2 / 2\Lambda$, where $\Delta Q_\alpha$ is the displacement along normal mode $\alpha$ between the ground and excited state geometries and $\Lambda = \sum_\alpha \Delta Q_\alpha^2 \omega_\alpha^2 / 2$ is the harmonic estimate of the molecular reorganisation energy. For the $\pi\to\pi^*$ state the geometry had to be constrained to be planar in the excited state geometry optimisation, to prevent optimisation to the \ce{S_1}-\ce{S_2} conical intersection and contributions from imaginary frequency modes are ignored. 
	The intramolecular reorganisation energy is set to be consistent with energy gap fluctuations obtained from the gas phase molecular dynamics/ESPF-DRF calculations\cite{fay_unraveling_2024,blumberger_recent_2015}
	\begin{align}
		\lambda_\mathrm{mol} = \frac{\ev{(\Delta E_{i\to f} - \ev{\Delta E_{i\to f}})^2}_\mathrm{gas}}{2k_\mathrm{B}T}.
	\end{align}
	The environment portion of the line-shape function $g_\mathrm{env}(t)$ is approximated using a classical high-temeprature Gaussian model
	\begin{align}
		g_\mathrm{env}(t) = -t^2 / (\hbar^2 k_\mathrm{B}T \lambda_\mathrm{env} )-i\lambda_\mathrm{env} t/\hbar
	\end{align}
	where the environment reorganisation energy is calculated from the energy gap fluctuations in the solution phase simulations
	\begin{align}
		\lambda_\mathrm{env} = \frac{\ev{(\Delta E_{i\to f} - \ev{\Delta E_{i\to f}})^2}_\mathrm{sol}}{2k_\mathrm{B}T} - \lambda_\mathrm{mol}.
	\end{align}
	Lastly the parameter $\epsilon$ is fitted based on the maximum in the static approximation spectrum,
	\begin{align}
		\epsilon = E_\mathrm{SA,max} - \lambda_\mathrm{mol} - \lambda_\mathrm{env}.
	\end{align}
	where $ E_\mathrm{SA,max} = \hbar\ \mathrm{argmax}_\omega A_{\mathrm{SA}}(\omega)$. The broadening parameter $\tau$ accounts for the finite lifetime of the excited state, as well as other broadening effects such as from the finite resolution of the spectrometer. This is treated as an empirical parameter fitted to the gas phase spectra. For the $n\to\pi^*$ transition this is set to \qty{100}{fs} and for the $\pi\to\pi^*$ transition it is set to \qty{75}{fs} (the position of the absorption maximum is not strongly dependent on this choice).
	
	This model for the absorption spectrum neglects a detailed treatment of non-Condon effects\cite{wiethorn_beyond_2023} (although it does account for them partially through the fitting the maximum of the static approximation spectrum), as well as potential Duschinsky rotation effects.\cite{miyazaki_singularity-free_2022} It also neglects the potential role of high frequency modes in the environment, and cross-correlation between molecular and environment degrees of freedom. Despite this the model provides an accurate description of the absorption line-shapes, and significantly more accurate than the simple static approximation spectra which completely misses the obvious asymmetry in the line shape.
	\vspace{-10pt}
	\subsection{Parameters for water models}
	\vspace{-10pt}
	The spectrum calculations use fixed charge, dipole-polarizable or fluctuating charge models for the \ce{H2O} polarizability. Several parameterizations of the models have been proposed so we test several of these in this work. The model parameters are summarised in tables \ref{tab-model-params-dip} and \ref{tab-model-params-fq}.\cite{nicoli_assessing_2022}
	\begin{table}[h!]
		\begin{tabular}{ccccc}
			% \hline
			\hline
			Model & $q_\mathrm{O}$ & $q_\mathrm{H}$ & $\alpha_{\mathrm{O}}$ & $\alpha_{\mathrm{H}}$  \\
			\hline 
			TIP3P & $-0.834$ & $+0.417$ & $0$ & $0$ \\
			Dipole 1 [Ref.~\onlinecite{jensen_discrete_2003}] & $-0.669$ & $+0.3345$ & $5.7494$ & $2.7929$ \\ 
			Dipole 2 [Ref.~\onlinecite{jensen_discrete_2003}] & $-0.669$ & $+0.3345$ & $9.718$ & $0$ \\
			\hline 
			% \hline
		\end{tabular}
		\caption{\ce{H2O} model parameters used in this work with a dipole-polarizability (or simple fixed point charge) model for the \ce{H2O} molecule. A Thole parameter of 2.1304 was used to damp the intramolecular induced-dipole interactions. Note that the Dipole 1 and Dipole 2 models here correspond to the DRF1 and DRF3 models in Ref.~\onlinecite{nicoli_assessing_2022}}\label{tab-model-params-dip}
	\end{table}
	\begin{table}[h!]
		\begin{tabular}{ccccc}
			% \hline 
			\hline
			Model & $\eta_\mathrm{O}$ & $\eta_\mathrm{H}$ & $\chi_\mathrm{O}$ & $\chi_\mathrm{H}$ \\
			\hline 
			FQ 1 [Ref.~\onlinecite{giovannini_effective_2019}] & 0.523700 & 0.537512 & 0.189194 & 0.012767 \\
			FQ 2 [Ref.~\onlinecite{rick_dynamical_1994}] & 0.584852 & 0.625010 & 0.116859 & 0 \\
			\hline
			% \hline
		\end{tabular}
		\caption{\ce{H2O} model parameters used in this work with a fluctuating charge model for the \ce{H2O} molecule. Note that the FQ 1 and FQ 2 models here correspond to the FQ3 and FQ1 models in Ref.~\onlinecite{nicoli_assessing_2022}.}\label{tab-model-params-fq}
	\end{table}
	\vspace{-10pt}
	\section{Gas phase acrolein spectra}\label{app-gas-spec}
	\vspace{-10pt}
	In order to test the spectrum calculation methods, we have used them to calculate the gas phase spectra.\cite{paulisse_vibronic_2000,aidas_performance_2008,lee_substituent_2007} In Fig.~\ref{fig-acrolein-vac} we show the calculated and experimental gas phase spectra.  For the $n\!\to\!\pi^*$ transition, Fig.~\ref{fig-acrolein-vac}A, the calculated spectra provide a reasonable estimate of the vibrational structure in the spectrum, but the peak position is captured fairly well, with an error of \qty{+0.06}{eV} for the uncorrected spectrum. Furthermore the 0-0 transition in the calculated spectrum is considerably blue-shifted relative to the experiment, and the experimental spectrum is significantly broader, which indicates that the TDDFT $\omega$B97X-D3/def2-TZVP method underestimates the intramolecular reorganisation energy. This is further evidences by the calculated \ce{C=O} stretching frequency, at \qty{1486}{cm^{-1}}, being considerably overestimated relative to experimental value at \qty{1173}{cm^{-1}}.\cite{paulisse_vibronic_2000} This leads to a systematic underestimation of the $n\to\pi^*$ absorption line-width. The agreement between calculated and experimental spectra is exceptionally good for the $\pi\!\to\!\pi^*$ transition, Fig.~\ref{fig-acrolein-vac}B, with an error of only \qty{-0.005}{eV} in the peak position. For both transitions the re-weighted spectra agree very well with the uncorrected spectra, with the uncorrected spectrum having an error of \qty{+0.022}{eV} and \qty{+0.033}{eV} in the peak positions of for the $n\!\to\!\pi^*$ and $\pi\!\to\!\pi^*$ transitions respectively. This shows that the force-field used for sampling is providing a good description of the acrolein intramolecular potential energy.
	
	\begin{figure}
		\centering
		\includegraphics[width=0.95\linewidth]{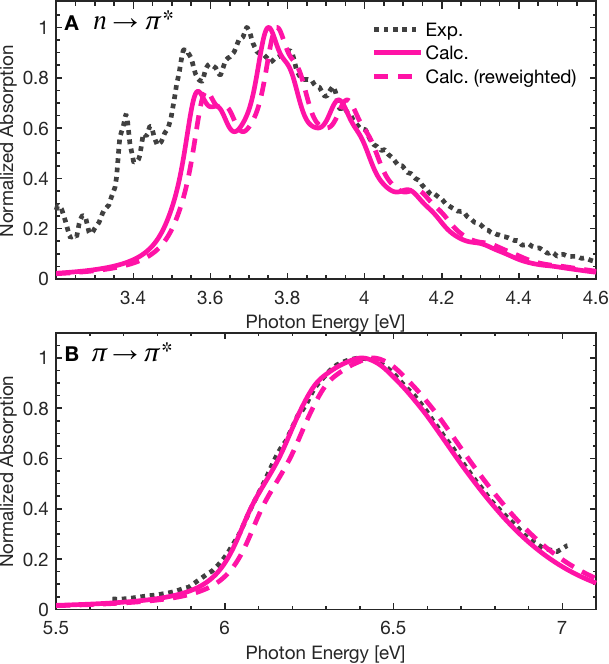}
		\caption{Normalized absorption spectra for acrolein in the region of the A) $n\!\to\!\pi^*$ transition and B) $\pi\!\to\!\pi^*$ transition. The uncorrected spectra (with MM force-field sampling of configurations)  are shown as solid lines and the re-weighted spectra are shown with solid dashed lines. Experimental spectra are shown as dotted lines. }
		\label{fig-acrolein-vac}
	\end{figure}

	\section*{References}
	\bibliography{refs}
	
\end{document}

\setlength{\abovedisplayskip}{4pt}
	\setlength{\belowdisplayskip}{4pt}
	
		{
		\makeatletter
		\def\frontmatter@thefootnote{%
			\altaffilletter@sw{\@fnsymbol}{\@fnsymbol}{\csname c@\@mpfn\endcsname}%
		}%
		\makeatother
	\title{Supporting Information to: Efficient polarizable QM/MM using the direct reaction field Hamiltonian with electrostatic potential fitted multipole operators}
	\author{Thomas P. Fay${}^1$}
	\email{tom.patrick.fay@gmail.com}
	\affiliation{${}^1$Aix Marseille Univ, CNRS, ICR, 13397 Marseille, France}
	%	\affiliation{	$\text{Corresponding author: tom.patrick.fay@gmail.com}$}
	\author{Nicolas Ferr\'e${}^1$}
	\affiliation{${}^1$Aix Marseille Univ, CNRS, ICR, 13397 Marseille, France}
	\author{Miquel Huix-Rotllant${}^1$}
	\affiliation{${}^1$Aix Marseille Univ, CNRS, ICR, 13397 Marseille, France}

	\maketitle
	\tableofcontents
	\section{Computational scaling of methods}
	
	Here we analyse the computational scaling of IEDRF and ESPF-DRF approaches. For hybrid-DFT and Hartree-Fock DRF methods, the most expensive term to evaluate in solving the KS/HF equations is the exchange term. In IEDRF this is given by\cite{humeniuk_multistate_2024}
	\begin{align}
		K_{\mu\nu}^{\mathrm{IEDRF}} &= -\sum_{\mu'=1}^{N_{\mathrm{AO}}} \sum_{p=1}^{3N_\MM}  \left(\sum_{\nu'=1}^{N_{\mathrm{AO}}}  D_{\mu'\nu'} {F}_{p,\mu\nu'}\right)\left(\sum_{q=1}^{3N_\MM}[\vb{K}_{\mathrm{D}}]_{pq} {F}_{q,\mu'\nu}\right) \\
		&=-\sum_{\mu'=1}^{N_{\mathrm{AO}}} \sum_{p=1}^{3N_\MM} [FD]_{p,\mu\mu'} [K_\mathrm{D}F]_{p,\mu'\nu} 
	\end{align}
	Evaluating $[FD]_{p,\mu\nu'}$ tensor scales as $\mathcal{O}(3 N_\MM N_{\AO}^3)$ and evaluating the $[K_\mathrm{D}F]_{p,\nu'\nu}$ tensor scales as $\mathcal{O}(9N_\MM^2 N_{\AO}^2)$, assuming $\vb{K}_\mathrm{D}\vb*{v}$ is evaluated iteratively and thus scales as $\mathcal{O}(9N_{\MM}^2)$. Once these terms are evaluated the final contraction scales as $\mathcal{O}(3 N_\MM N_{\AO}^3)$. In ESPF-DRF this term is evaluated as
	\begin{align}
		K_{\mu\nu}^{\mathrm{ESPF-DRF}} &=-\sum_{\mu'=1}^{N_{\mathrm{AO}}} \sum_{a=1}^{N_\mathcal{Q}}   \left( \sum_{\nu'=1}^{N_{\mathrm{AO}}} D_{\mu' \nu'} \mathcal{Q}_{a,\mu\nu'} \right) \left(\sum_{b=1}^{N_\mathcal{Q}} U_{\QM}^{ab} \mathcal{Q}_{b,\nu\mu'}\right) \\
		&=-\sum_{\mu'=1}^{N_{\mathrm{AO}}} \sum_{a=1}^{N_\mathcal{Q}}  [D\mathcal{Q}]_{a,\mu\mu'} [U_{\QM}\mathcal{Q}]_{a,\mu'\nu}
	\end{align}
	Similar to in IEDRF calculating the $[D\mathcal{Q}]_{a,\mu\nu'}$ tensor scales as $\mathcal{O}(N_\mathcal{Q} N_{\AO}^3)$ and $[U_{\QM}\mathcal{Q}]_{a,\nu'\nu}$ scales as $\mathcal{O}(N_\mathcal{Q}^2 N_{\AO}^2)$, and for large MM systems evaluating $U_{\QM}^{ab}$ scales as $\mathcal{O}(9N_{\MM}^2 N_{\mathcal{Q}})$. 
	
	Comparing the two methods, if the final contraction dominates computation for IEDRF the method scales as $\mathcal{O}(3N_{\MM}N_{\AO}^3)$ whereas ESPF-DRF scales as $\mathcal{O}(N_{\mathcal{Q}} N_{\AO}^3)$, so ESPF-DRF offers a speed-up of $3N_{\MM}/N_{\mathcal{Q}}$. In the limit where the MM systems is very large evaluation of $[K_\mathrm{D}F]_{p,\nu'\nu}$ dominates in IEDRF and $U_{\QM}^{ab}$ dominates for ESPF-DRF, meaning that the methods scale as $\mathcal{O}(9N_{\MM}^2 N_{\AO}^2)$ and $\mathcal{O}(9N_{\MM}^2 N_{\mathcal{Q}})$ respectively, so ESPF-DRF in this case offers a speed-up of $N_{\AO}^2/N_{\mathcal{Q}}$.
	
	A similar analysis can be applied to the one-electron term, which for IEDRF involves the same $[K_\mathrm{D}F]_{p,\nu'\nu}$ tensor, where it can be found that the final contraction in the dominant electron self-interaction term scales as $\mathcal{O}(N_{\MM}N_{\AO}^3)$, whereas for ESPF-DRF it scales as $\mathcal{O}(N_{\mathcal{Q}}N_{\AO}^3)$, so again for when $N_{\AO}^3$ terms dominate, the ESPF-DRF method offers a speed-up of $3N_{\MM}/N_{\mathcal{Q}}$ whereas when the $N_{\MM}^2$ term dominates it offers a speed-up of $N_{\AO}^2/N_{\mathcal{Q}}$, the same as the evaluation of the exchange term.
	
	To put some nubers to these speed-ups for a real system, in the acrolein example in the main text each snapshot contains approximately $2000$ polarizable sites, and with 8 atoms in the QM region $N_\mathcal{Q} = 32$, and so the final contraction in evaluation of the exchange matrix is sped up by a factor of $3N_\MM / N_\mathcal{Q} \approx 180$. The speed-up of the evaluation of the $[K_\mathrm{D}F]_{p,\mu\nu}$ tensor compared to the evaluation of the analogous $[U_{\QM}\mathcal{Q}]_{a,\mu\nu}$ for acrolein with the def2-TZVP basis is about $N_\AO^2 / 2 N_\mathcal{Q} \approx 680$. 
	
	For the mean-field approach, there is no correction to the exchange matrix to evaluate, instead there is only the evaluation of a corrected Coulomb matrix
	\begin{align}
		J^{\mathrm{MF}}_{\mu\nu} &= -\sum_{p=1}^{3 N_\MM} F_{p,\mu\nu} \mu^{\QM}_{p}\\
		\mu^{\QM}_{p} &= \sum_{q=1}^{3 N_\MM} [\vb{K}_\mathrm{D}]_{pq}\sum_{\mu'=1}^{N_\AO} \sum_{\nu'=1}^{N_\AO} D_{\mu'\nu'}F_{q,\mu'\nu'}
	\end{align}
	the final contraction scales as $\mathcal{O}(3N_\MM N_\AO^2 )$, whilst the evaluation of $\mu^{\QM}_{p}$ scales as $\mathcal{O}(9N_{\MM}^2)$ for the $[\vb{K}_\mathrm{D}]_{pq} $ multiplication step, and $\mathcal{O}(3N_\MM N_{\AO}^2)$ for the evaluation of expection value of the fields $\Tr[\vb{D}\vb{F}_p]$ at each MM site. So the mean-field method is $\mathcal{O}(N_{\AO})$ times faster than the IEDRF approach for typical calculations. The ESPF-MF approach, where the ESPF method is used to evaluate the polarization energy (as is done in the main text), also accelerates the evaluation of the MF Coulomb matrix,
	\begin{align}
		J^{\mathrm{ESPF-MF}}_{\mu\nu} &= -\sum_{a=1}^{N_\mathcal{Q}} \mathcal{Q}_{a,\mu\nu} \phi^{\QM}_{a}\\
		\phi^{\QM}_{a} &= \sum_{b=1}^{N_\mathcal{Q}} U_{\QM}^{ab}\sum_{\mu'=1}^{N_\AO} \sum_{\nu'=1}^{N_\AO} D_{\mu'\nu'}\mathcal{Q}_{b,\mu'\nu'}
	\end{align}
	The final contraction here is now $\mathcal{O}(N_\mathcal{Q} N_\AO^2)$ and evaluation of the expectation value of the multipoles $\Tr[\vb{D}\vb*{\mathcal{Q}}_a]$ is $\mathcal{O}(N_\mathcal{Q} N_{\AO}^2)$, and evaluation of $U_{\QM}^{ab}$ is $\mathcal{O}(9 N_\MM^2 N_\mathcal{Q})$ but it is only performed once, rather than at every step of the self-consistent field cycle. It is in principle possible to develop an algorithm that will only require $\min(N_\mathcal{Q},N_{\mathrm{SCF}})$ solutions to the induction equations for ESPF-MF by storing solutions of the induction equations from each step a complete linearly independent set of solutions is formed, where then $U_{\QM}^{ab}$ can be formed from this set of solutions. Our implementation of ESPF-MF only uses the simpler algorithm above.
	
	\section{Comparing corrected and uncorrected ESPF approaches}
	
	Here we compare the ESPF-DRF using the corrected ESPF operators (as done in the main text), with uncorrected operators, where the exact total charge and dipole operator constraints are not enforced. This is done for the models in Figs. 1--3 in the main text. These are given in Figs.~\ref{espf1}, \ref{espf2} and \ref{espf3}. With the exception of the \ce{HF+He} curve in the \ce{S_1} state the corrected and uncorrected curves give identical results.
	
	\begin{figure}[h!]
		\includegraphics[width=0.5\linewidth]{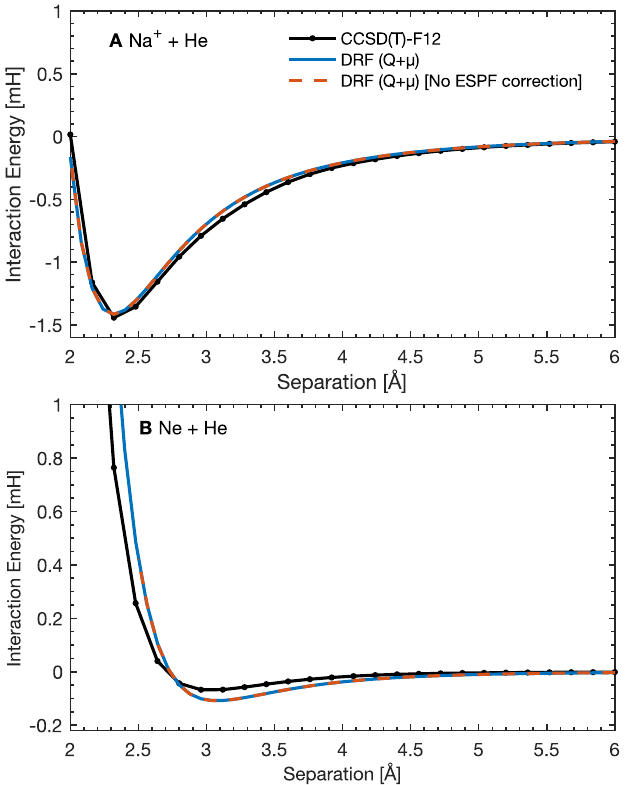}
		\caption{A) \ce{Na+ + He} and B) \ce{Ne + He} interaction energy curves with corrected (same as main text) and uncorrected ESPF operators.}\label{espf1}
	\end{figure}
	\begin{figure}[h!]
		\includegraphics[width=0.5\linewidth]{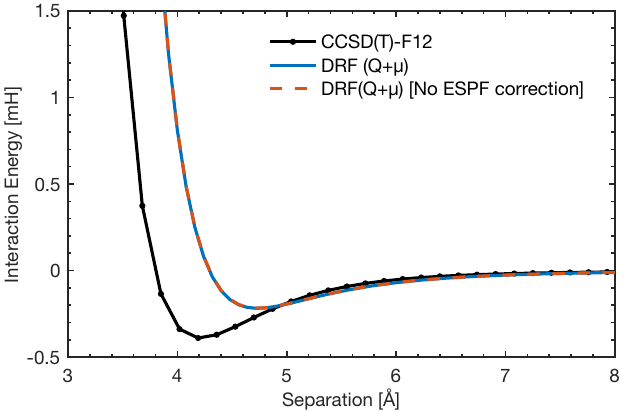}
		\caption{\ce{CH4 + Ar}  interaction energy curves with corrected (same as main text) and uncorrected ESPF operators.}\label{espf2}
	\end{figure}
	\begin{figure}[h!]
		\includegraphics[width=0.5\linewidth]{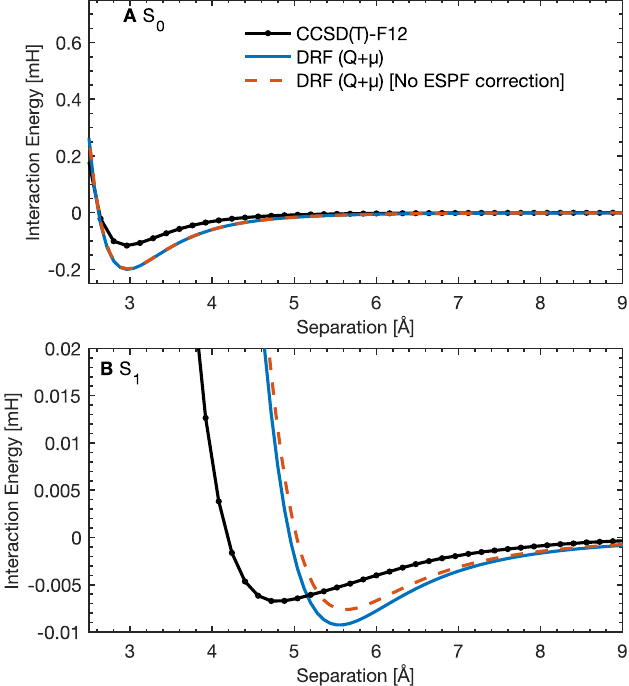}
		\caption{\ce{HF + Ar} interaction energy curves with corrected (same as main text) and uncorrected ESPF operators for A) the \ce{S_0} state and B) the \ce{S_1} state.}\label{espf3}
	\end{figure}
	
	\section{The choice of damping in ESPF-DRF}
	
	In all of the results shown in the main text the covalent radius based damping scheme is used. Here we explore the effect of the choice of damping by also exploring setting the damping radii to a universal constant $R^{\mathrm{damp}}_{A,B} = R_{\mathrm{damp}}$ between $ R_\mathrm{damp} = 1 \ \mathrm{a}_0$ and $ R_\mathrm{damp} = 3 \ \mathrm{a}_0$. These results are shown in Figs.~\ref{damp1}, \ref{damp2} and \ref{damp3}. With th exception of the $\ce{Na+ +  He}$ case in Fig.~\ref{damp1}A, $R_\mathrm{damp}$ in the range 1 to 2 $\mathrm{a}_0$ has very little effect on the interaction curves. In the case of \ce{Na+ +  He} 
	the interaction is a very strong ion-induced dipole interaction involving two relatively small atoms, with an equilibrium separation of about 2.3 \AAA, so it is unsurprising that this case is the most sensitive to the choice of damping parameter. These results suggest that in cases involving small ions in QM region, care should be taken to check the applicability of the covalent radius based damping scheme, and if necessary tune the damping parameter. In this case the covalent radius of neutral Na (1.66 \AAA) is not appropriate (with a [Ne]3s${}^1$ electron configuration) and instead the effective ionic radius is chosen 1.02 \AAA.
	
	\begin{figure}[h!]
		\includegraphics[width=0.5\linewidth]{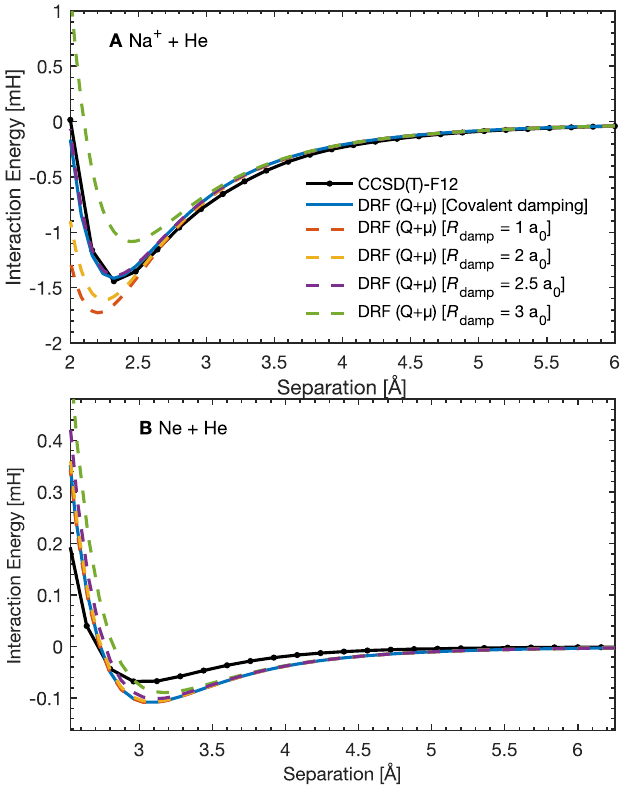}
		\caption{A) \ce{Na+ + He} and B) \ce{Ne + He} interaction energy curves with corrected (same as main text) and uncorrected ESPF operators.}\label{damp1}
	\end{figure}
	\begin{figure}[h!]
		\includegraphics[width=0.5\linewidth]{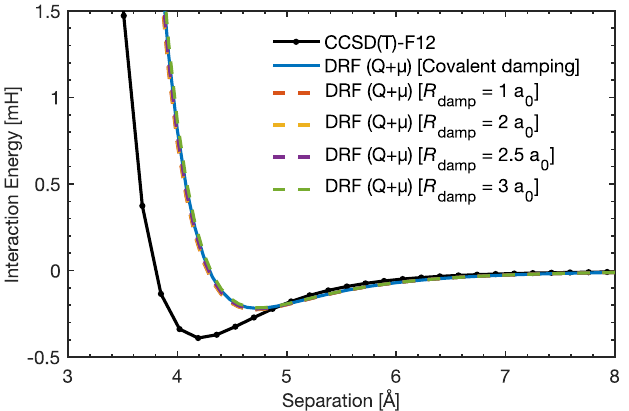}
		\caption{\ce{CH4 + Ar}  interaction energy curves with corrected (same as main text) and uncorrected ESPF operators.}\label{damp2}
	\end{figure}
	\begin{figure}[h!]
		\includegraphics[width=0.5\linewidth]{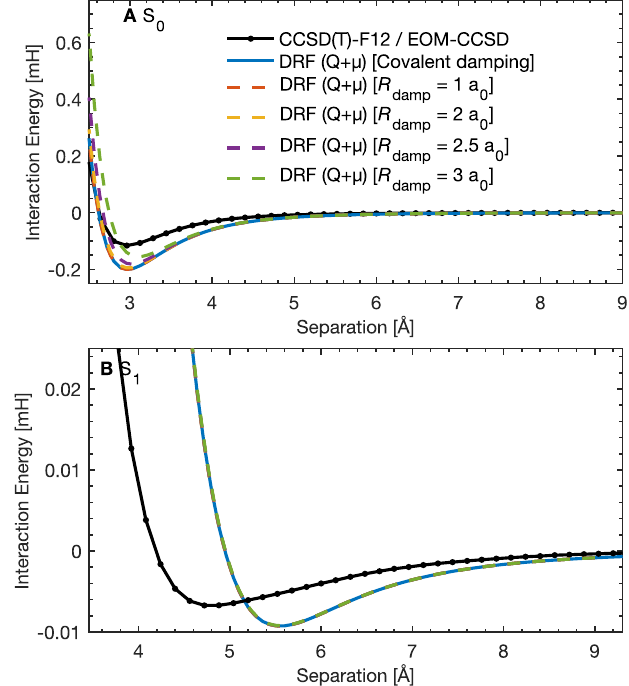}
		\caption{\ce{HF + He} interaction energy curves with corrected (same as main text) and uncorrected ESPF operators for A) the \ce{S_0} state and B) the \ce{S_1} state.}\label{damp3}
	\end{figure}

	\section*{References}
	\bibliography{sirefs.bib}